\documentclass[twocolumn]{aastex63}

\usepackage[utf8]{inputenc}
\usepackage{rotating}
\usepackage{tablefootnote}
\usepackage{natbib}
\usepackage{amsmath}
\usepackage{threeparttable}
\usepackage{xspace}
\usepackage{graphicx}
\usepackage{longtable}
\usepackage{multirow}
\usepackage{hyperref}

\newcommand{\rearth}{R$_\oplus$\xspace}

\begin{document}

\title{A Uniform Search for Nearby Planetary Companions to Hot Jupiters in TESS Data Reveals Hot Jupiters are Still Lonely}

\author[0000-0001-5084-4269]{Benjamin J. Hord}
\affiliation{Department of Astronomy, University of Maryland, College Park, MD 20742, USA}
\affiliation{NASA Goddard Space Flight Center, 8800 Greenbelt Road, Greenbelt, MD 20771, USA}
\affiliation{GSFC Sellers Exoplanet Environments Collaboration}

\author[0000-0001-8020-7121]{Knicole D. Col\'{o}n}
\affiliation{NASA Goddard Space Flight Center, 8800 Greenbelt Road, Greenbelt, MD 20771, USA}
\affiliation{GSFC Sellers Exoplanet Environments Collaboration}

\author[0000-0001-9786-1031]{Veselin Kostov}
\affiliation{NASA Goddard Space Flight Center, 8800 Greenbelt Road, Greenbelt, MD 20771, USA}
\affiliation{GSFC Sellers Exoplanet Environments Collaboration}

\author[0000-0001-5379-4295]{Brianna Galgano}
\affiliation{Department of Physics and Astronomy, Johns Hopkins University, 3400 North Charles Street, Baltimore, MD 21218, USA}
\affiliation{NASA Goddard Space Flight Center, 8800 Greenbelt Road, Greenbelt, MD 20771, USA}

%\author{others!}

%\author{TESS Architects}

\author[0000-0003-2058-6662]{George~R.~Ricker}
\affiliation{Department of Physics and Kavli Institute for Astrophysics and Space Research, Massachusetts Institute of Technology, Cambridge, MA 02139,
USA}

\author[0000-0001-6763-6562]{Roland~Vanderspek}
\affiliation{Department of Physics and Kavli Institute for Astrophysics and Space Research, Massachusetts Institute of Technology, Cambridge, MA 02139,
USA}

\author[0000-0002-6892-6948]{S.~Seager}
\affiliation{Department of Physics and Kavli Institute for Astrophysics and Space Research, Massachusetts Institute of Technology, Cambridge, MA 02139,
USA}
\affiliation{Department of Earth, Atmospheric and Planetary Sciences, Massachusetts Institute of Technology, Cambridge, MA 02139, USA}
\affiliation{Department of Aeronautics and Astronautics, MIT, 77 Massachusetts Avenue, Cambridge, MA 02139, USA}

\author[0000-0002-4265-047X]{Joshua~N.~Winn}
\affiliation{Department of Astrophysical Sciences, Princeton University, 4 Ivy Lane, Princeton, NJ 08544, USA}

\author[0000-0002-4715-9460]{Jon~M.~Jenkins}
\affiliation{NASA Ames Research Center, Moffett Field, CA 94035, USA}

%\author{TESS Contributed Authors}

\author[0000-0001-7139-2724]{Thomas~Barclay}
\affiliation{NASA Goddard Space Flight Center, 8800 Greenbelt Road, Greenbelt, MD 20771, USA}
\affiliation{University of Maryland, Baltimore County, 1000 Hilltop Circle, Baltimore, MD 21250, USA}

\author[0000-0003-1963-9616]{Douglas~A.~Caldwell}
\affiliation{NASA Ames Research Center, Moffett Field, CA 94035, USA}
\affiliation{SETI Institute, Mountain View, CA 94043, USA}

%\author{John~P.~Doty}
%\affiliation{Noqsi Aerospace Ltd., 15 Blanchard Avenue, Billerica, MA 01821, USA}

\author[0000-0002-2482-0180]{Zahra~Essack}
\affiliation{Department of Earth, Atmospheric and Planetary Sciences, Massachusetts Institute of Technology, Cambridge, MA 02139, USA}
\affiliation{Kavli Institute for Astrophysics and Space Research, Massachusetts Institute of Technology, Cambridge, MA 02139, USA}

\author[0000-0002-9113-7162]{Michael~Fausnaugh}
\affiliation{Department of Physics and Kavli Institute for Astrophysics and Space Research, Massachusetts Institute of Technology, Cambridge, MA 02139, USA}

\author[0000-0002-5169-9427]{Natalia~M.~Guerrero}
\affiliation{Department of Physics and Kavli Institute for Astrophysics and Space Research, Massachusetts Institute of
Technology, Cambridge, MA 02139, USA}

%\author{Richard~C.~Kidwell~Jr.}
%\affiliation{Space Telescope Science Institute, Baltimore, MD, USA}

\author[0000-0002-5402-9613]{Bill~Wohler}
\affiliation{NASA Ames Research Center, Moffett Field, CA 94035, USA}
\affiliation{SETI Institute, Mountain View, CA 94043, USA}
\correspondingauthor{Benjamin J. Hord}
\email{benhord@astro.umd.edu}

\begin{abstract}
    We present the results of a uniform search for additional planets around all stars with confirmed hot Jupiters observed by the Transiting Exoplanet Survey Satellite (TESS) in its Cycle 1 survey of the southern ecliptic hemisphere. Our search comprises 184 total planetary systems with confirmed hot Jupiters with $R_{p}$ $>$ 8\rearth and orbital period $<$10 days. The Transit Least Squares (TLS) algorithm was utilized to search for periodic signals that may have been missed by other planet search pipelines. While we recovered 169 of these confirmed hot Jupiters, our search yielded no new statistically-validated planetary candidates in the parameter space searched ($P <$ 14 days). A lack of planet candidates nearby hot Jupiters in the TESS data supports results from previous transit searches of each individual system, now down to the photometric precision of TESS. This is consistent with expectations from a high eccentricity migration formation scenario, but additional formation indicators are needed for definitive confirmation. We injected transit signals into the light curves of the hot Jupiter sample to probe the pipeline's sensitivity to the target parameter space, finding a dependence proportional to $R_{p}^{2.32}P^{-0.88}$ for planets within 0.3$\leq$$R_{p}$$\leq$4 \rearth and 1$\leq$$P$$\leq$14 days. A statistical analysis accounting for this sensitivity provides a median and $90\%$ confidence interval of $7.3\substack{+15.2 \\ -7.3}\%$ for the rate of hot Jupiters with nearby companions in this target parameter space. This study demonstrates how TESS uniquely enables comprehensive searches for nearby planetary companions to nearly all the known hot Jupiters.
\end{abstract}


\section{Introduction} \label{sec:intro}

Hot Jupiters (HJs) were among the most surprising class of planets discovered by the first exoplanet surveys. Both the first exoplanet discovered around a main sequence star \citep{mayor1995jupiter} and the first known transiting exoplanet \citep{charbonneau1999detection, henry1999transiting} were HJs. With radii of $R_{p}$ $>$ 8 \rearth and orbital periods of $P$ $<$ 10 days \citep{wang2015occurrence, winn2010hot, garhart2020statistical, huang2016warm}, HJs are unlike any planet in the Solar System. Many scenarios have been put forth to explain the existence of HJs, such as disk migration \citep{lin1996orbital}, high-eccentricity migration (HEM) \citep{rasio1996dynamical}, \textit{in situ} formation \citep{mayor1995jupiter}, and many others; however, none of these formation mechanisms can explain the observed properties of every HJ system \citep{dawson2018origins}.

Notably, HJs are often the only detected planet in their systems out to an orbital period of $\sim$200 days \citep{knutson2014friends, endl2014kepler, steffen2012kepler}. Previous searches for companions to HJs using ground- or space-based data have returned only three known systems (WASP-47, Kepler-730, and TOI-1130) with a HJ and nearby companion planets, out of the many hundreds of currently confirmed HJ systems \citep{becker2015wasp, zhu2018kepler, canas2019kepler, huang2020tess}. The radii of these companions are 3.58 \rearth, 1.80 \rearth, 1.57 \rearth, and 3.65 \rearth for WASP-47 d, WASP-47 e, Kepler-730 c, and TOI-1130 b, respectively, making them all smaller than Neptune. Combined with their short ($\leq$10 days) orbital periods, the transit signals from this class of small planets could be easily missed by planet search pipelines or in noisy data.

The apparent lack of nearby companions in the vast majority of HJ systems supports the idea that HJs form beyond the ice line and migrate inwards via HEM, which would destabilize the orbits of any shorter-period planets in the system \citep{mustill2015destruction}. This may not be the full story however, since not only are some systems known to have companion planets, but statistical work based on photometric observations suggests that some fraction of HJ systems could have formed via methods other than HEM based on the lack of eccentric proto-HJs observed \citep{dawson2014photoeccentric}, although it is difficult to rule out HEM entirely for these systems.

This leaves the formation mechanism for many HJs largely a mystery with multiple possibilities for a given individual system. It is possible that the three unique HJs with companion planets named above formed via a different mechanism from many of the other HJs or that they may simply be rare variants of HJs. Comprehensive searches for companion planets to HJs could reveal additional nearby companions to HJs or support previous findings as to the ``loneliness'' of HJs.

All-sky transit surveys conducted with ground-based telescopes typically do not reach the photometric precision needed to identify shallow transit signals of small, nearby companions to hot Jupiters \citep{pollacco2006wasp, bakos2004wide, pepper2007kilodegree}. The Kepler and K2 missions led to the discovery of two of three known HJ systems with nearby companion planets \citep{becker2015wasp, canas2019kepler}, but both missions surveyed only a small fraction of the sky \citep{borucki2010kepler, howell2014k2}. TESS has enabled nearly-full-sky coverage observations with space-based photometric precision optimized for the discovery of exoplanets around bright stars and presents an excellent opportunity to finally conduct a uniform search for these additional, closely-orbiting planets. In fact, TESS has already begun to demonstrate its usefulness in the search for HJs with companions since the most recent of the systems with companions near an HJ (TOI-1130) was discovered by TESS \citep{huang2020tess}.

Much of the TESS data, including most of the HJ systems, are searched by the TESS Science Processing Operations Center (SPOC) pipeline \citep{Jenkins2016} and/or the MIT Quicklook Pipeline (QLP) \citep{huang2020photometry, huang2020photometry2} prior to each data release to the general public. After extracting simple aperture photometric light curves from the calibrated pixel data, the SPOC pipeline identifies and corrects instrumental systematic errors and flags bad data with the Presearch Data Conditioning (PDC) module \citep{smith2012kepler, stumpe2012kepler, stumpe2014multiscale}. The SPOC then searches through the resulting PDC$\_$SAP light curves using a wavelet-based, adaptive matched filter algorithm to detect signatures of potential transiting planets \citep{2002ApJ...575..493J, 2010SPIE.7740E..0DJ, jenkins2017kepler}. Limb-darkened transit models are fitted to each of these “threshold crossing events” \citep{Li:DVmodelFit2019}, and are then subjected to a suite of diagnostic tests by the Data Validation (DV) module to help adjudicate the planetary nature of each signal \citep{Twicken:DVdiagnostics2018}. The TESS Science Office reviews the DV reports and diagnostics and promotes and releases compelling cases as TESS Objects of Interest (TOI) for follow up and characterization. Parallel to the SPOC, the QLP extracts its own light curves from the TESS data and searches all targets down to a $Tmag$ of 13.5 using a Box Least Squares (BLS) search algorithm \citep{guerrero2021tess, huang2020photometry, huang2020photometry2}.

Here, we implement in our pipeline the recently published Transit Least Squares (TLS) algorithm \citep{2019A&A...623A..39H}. Unlike BLS, TLS utilizes a realistic transit model derived from fitting 2,346 known exoplanet light curves that takes into account many of the physical parameters of the system, such as host star mass, radius, and limb darkening parameters. Most notably, TLS provides a 17$\%$ increase in detection efficiency for the signals produced by smaller planets over BLS \citep{2019A&A...623A..39H}, although some work has shown that realistic transit shapes only provide as low as a $\sim$3$\%$ sensitivity increase if BLS is sufficiently well-sampled \citep[e.g.,][]{jenkins1996matched}. This higher detection efficiency provides the opportunity to recover a greater proportion of planets with smaller radii than BLS. It also opens up a parameter search space complementary to the QLP and SPOC pipeline and increases the significance of signals considered marginal by BLS. It should also be noted that there has, as of yet, been no direct comparison between TLS and the search conducted by the SPOC pipeline in terms of sensitivity.

In this paper, we present a uniform search for nearby transiting companions to all confirmed HJs observed in the southern ecliptic hemisphere by TESS during its Cycle 1 observations that is meant to be independent of the SPOC and MIT pipelines. It is well-documented that different search pipelines often have different recovery rates and result in transit detections in different parts of the planetary parameter space \citep[e.g.,][]{kostov2019a, kruse2019detection}. Therefore, we searched for transit signals assuming no prior knowledge of SPOC pipeline or QLP detections with the aim of providing a separate search that utilized different search methods. Our search is uniform for signals with periods $<$ 14 days ($\sim$half the duration of a TESS sector) and potential planets with $R_{p}$ $\leq$ 4 \rearth around HJ-bearing systems with host stars of 7.3 $\leq$ $Tmag$ $\leq$ 19.9 in the TESS Cycle 1 data.

\begin{figure*}
    \centering
    \includegraphics[width=1.0\textwidth]{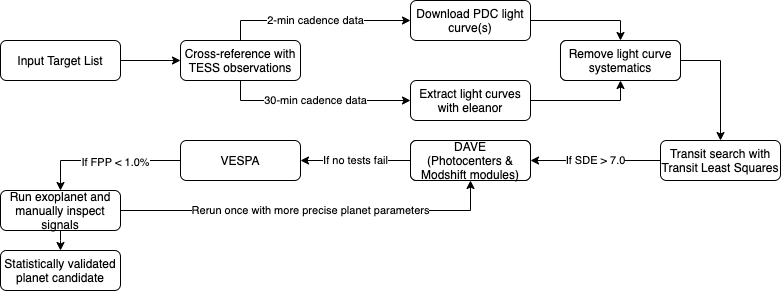}
    \caption{A schematic outline of the processing pipeline used in this study. Illustrated are the major steps in the search for additional transit signals in each TESS light curve as well as conditions which, if met, caused a target to advance to the next stage of analysis. In the final loop of the pipeline after the transit search, signals underwent an initial round of vetting and validation with \texttt{DAVE} (Discovery and Validation of Exoplanets) and \texttt{vespa} before being modeled by \texttt{exoplanet}. After the signals were modeled and more precise transit parameters were acquired, the signals were run through \texttt{DAVE} and \texttt{vespa} once more for a final round of vetting and validation using these more precise parameters.}
    \label{fig:pipeline_flowchart}
\end{figure*}

Section \ref{sec:observations} describes how the initial target list of HJs was compiled and the TESS time series data products utilized by this study. Section \ref{sec:pipeline} outlines the search methods used to find periodic transit-like signals in each light curve as well as the tools used to vet and validate each of these new signals. Section \ref{sec:exoplanet} contains a description of how the precise planet and orbital parameters were modeled for each system. Section \ref{sec:injection} outlines the statistical analysis conducted to estimate the rate of nearby companions to HJs. Section \ref{sec:discussion} discusses implications of our findings with regards to HJ formation, and we present a summary of our conclusions in Section \ref{sec:conclusion}.

\section{Target Selection and Data Acquisition} \label{sec:observations}

For the purposes of this study and in order to encompass a wide data set, a planet was considered a hot Jupiter if it had an orbital period of $P<10$ days and a radius of $R_p$ $>$ 8\rearth $\approx$ 0.71 $R_{J}$ \citep{wang2015occurrence, winn2010hot, garhart2020statistical, huang2016warm}. The NASA Exoplanet Archive\footnote{https://exoplanetarchive.ipac.caltech.edu/} was queried on January 6, 2020 with these parameters, resulting in a dataset comprised of 437 confirmed HJs. The Right Ascension, Declination, Common Name, Orbital Period, and Radius of each planet were downloaded for use in our analysis pipeline.

This study was restricted to the first year of the TESS prime mission, which covered the southern ecliptic hemisphere and corresponds to TESS Sectors 1-13. This complements observations being collected currently in the TESS extended mission, where TESS is revisiting the southern ecliptic hemisphere between July 2020 and June 2021. We used the \texttt{Tesscut} module of the \texttt{astroquery.mast} Python package \citep{2019ascl.soft05007B} to determine that 183 of the total 437 HJs in the dataset were observed in the first 13 sectors of the TESS prime mission. The TOI-1130 system was added in after the creation of this HJ data set due to the discovery of a nearby companion to the HJ in the system. This brought the total data set up to 184 HJs spanning 0.77 days to 9.62 days in orbital period and 9.41 \rearth to 21.41 \rearth in radius.

The host stars for these systems have effective temperatures ranging from 3749 K to 9364 K and 97$\%$ of the targets fall within the main sequence F, G, and K type stellar classifications. The 5 targets that do not fall within the F, G, and K stellar classifications are all classified as main sequence A type stars. There is only one young star in the sample - DS Tuc. Additional information on each of these targets can be found in the TESS Input Catalog Version 8 (TIC, \cite{stassun2019revised}) and all subsequent analysis uses stellar values gathered from the TIC.

In the prime mission, the four TESS cameras captured a Full Frame Image (FFI) of each $\sim$27-long day observation sector every 30 minutes while $\sim$200,000 pre-determined targets had a smaller image captured at a cadence of 2 minutes, providing superior data quality for determining transit parameters. All TESS data are calibrated by the SPOC at NASA Ames Research Center. The targets observed at 2-minute cadence also have Pre-search Data Conditioning (PDC) light curves which are systematic error-corrected using an optimal photometric aperture \citep{smith2012kepler, stumpe2014multiscale, jenkins2016tess}. These light curves have also been corrected for instrumental signals and contaminating light from nearby stars. There were 126 of the 184 HJs observed in the first 13 sectors of the TESS mission that were observed at 2-minute cadence and had PDC light curves generated in addition to the longer 30-minute cadence data extracted from the TESS FFIs. Both cadences for each target were used in this analysis.

The 2-minute PDC light curves were downloaded from the Mikulski Archive for Space Telescopes (MAST) while the 30-minute light curves were extracted from the TESS FFIs using the \texttt{eleanor} Python package, an open-source tool to produce light curves for objects \citep{feinstein2019eleanor}. In short, \texttt{eleanor} generates light curves for various combinations of pre-set apertures to determine which aperture minimizes the combined differential photometric precision (CDPP) for data binned to a cadence of 1 hour.

\section{Transit Search, Vetting, and Validation} \label{sec:pipeline}

After each light curve is extracted, Transit Least Squares (TLS) is used to search for periodic, transit-like signals. Significant signals are then passed through \texttt{DAVE} (Discovery and Validation of Exoplanets) for vetting and \texttt{vespa} for validation.

\subsection{Periodic Signal Search} \label{ssec:signal_search}

We used the methods presented in \citet{heller2019transit} as a guide to prepare the TESS light curves for our planet search and for implementing the TLS algorithm. The light curves were iteratively clipped of outliers $>$3$\sigma$ and detrended using \texttt{lightkurve}'s built-in \texttt{flatten} method \citep{2018ascl.soft12013L} which applies a Savitzky-Golay filter to remove low frequency trends in the light curve by fitting successive sub-sets of adjacent data points with a low-degree polynomial. A window length of $\sim$0.5 days was selected as it compromises between a short enough window to remove stellar variability while still keeping transits intact since the transit duration for all HJs in the sample are $\leq$9 hours. Known HJ transits were masked during this filtering using the orbital periods and transit epoch queried from the MAST. TLS was then run on each processed light curve using the default settings and input stellar parameters from the TIC. We considered a periodic signal to be significant if its signal detection efficiency (SDE) $>$7.0, which corresponds to a false alarm probability (FAP) that the signal is a result of statistical fluctuations of $<$1$\%$ based on 10000 transit injection simulations performed by \cite{2019A&A...623A..39H} using the TLS algorithm on simulated Kepler data with a time baseline of 3 years.

Both the 30-minute cadence and 2-minute cadence light curves were run separately through the transit search as two independent searches of all available data using identical methods for both. Additionally, if a target had more than one sector of data, a transit search was run on each sector individually as well as on the full, combined light curve. This was done to mitigate sector-dependent systematic effects (e.g., impacts of scattered light). For each target, only signals with a period of up to half the total observation length were considered to ensure at least two transits were contained within the observation. Because many targets were only observed for a single TESS sector ($\sim$28 days), we cannot rule out the existence of transiting planets beyond a period of 14 days and are most confident for signals with $P <$ 14 days.

Each target was searched with both the TLS ``default" shape (more U-shaped) as well as the ``grazing" shape (more V-shaped) to maximize the SDE of any possible signals found. TLS was run iteratively for each shape and sector/cadence combination until the signal recovered did not meet the SDE $>$7.0 criterion. For each iteration, previous significant signals were masked out of the light curve for subsequent runs. No more than 2 significant signals in addition to the HJ were found for any target.

\begin{figure*}
    \centering
    \includegraphics[width=0.8\textwidth]{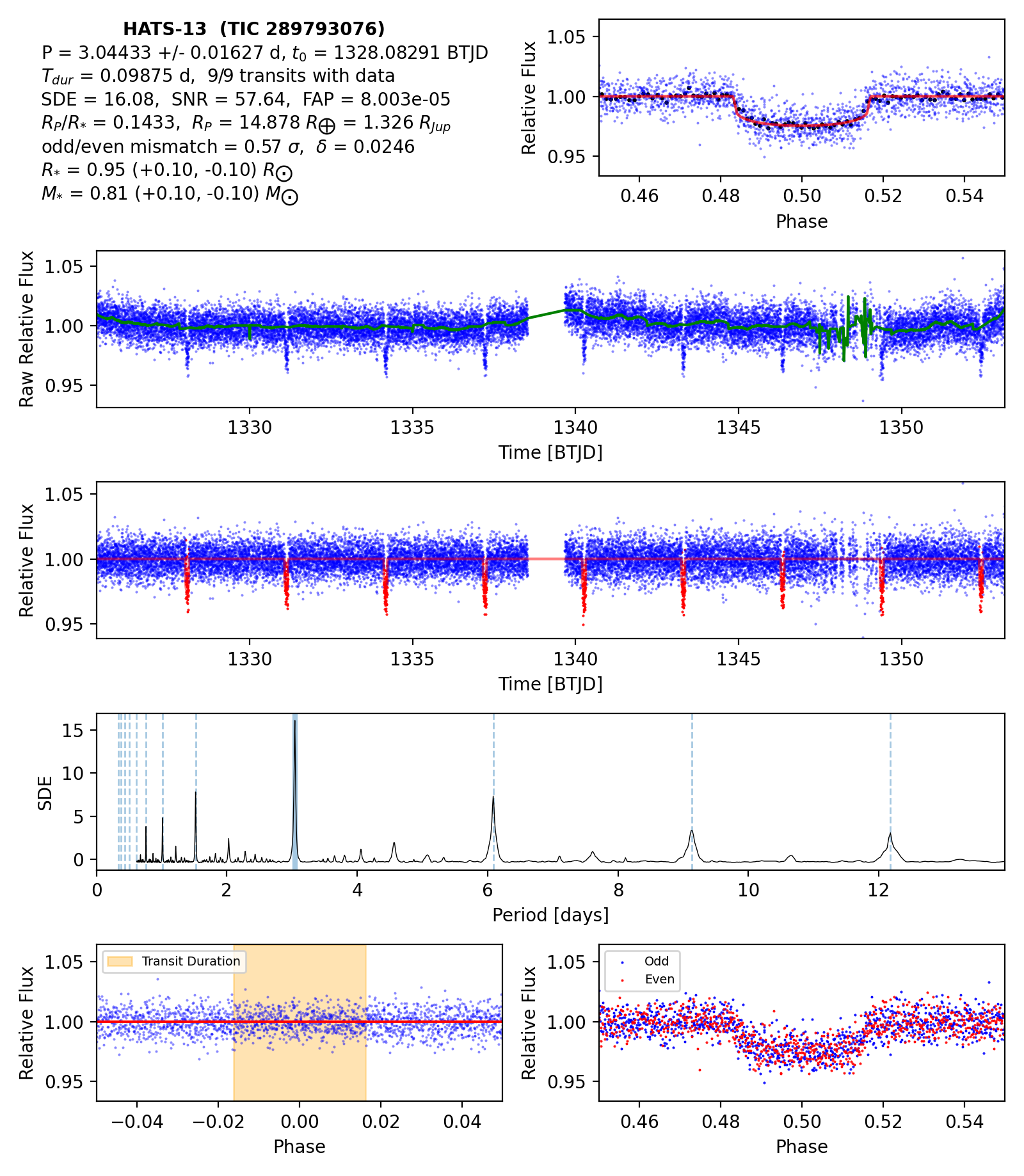}
    \caption{Example of the information printed on the vetting sheet for a single TLS iteration (i.e. detection of a single periodic signal). From top to bottom: orbital and system information for the signal, light curve phase-folded to period of the detected signal, raw light curve with removed trend overlaid, flattened light curve with transit model from TLS overlaid and in-transit points highlighted, TLS periodogram with strongest signal and integer multiples highlighted, light curve phase folded viewed a half phase apart from the transit with transit duration highlighted to search for secondary eclipses, odd/even transit comparison.}
    \label{fig:vet_sheet}
\end{figure*}

For each TLS iteration, key diagnostic parameters were output to a ``vetting sheet" which allowed for quick visual vetting of signals to identify any as obvious noise. The vetting sheet contains information such as best-fit orbital parameters, the phase-folded light curve, an odd-even transit comparison, the periodogram, a half-period check, and other items that are useful in determining whether a signal could be real or spurious. See Figure \ref{fig:vet_sheet} for an example of the vetting sheet used.

The TLS search of the 126 HJ systems in the 2-minute cadence data yielded 242 non-HJ signals with SDE $>$ 7.0. There were zero new, non-HJ signals in the 30-minute FFI light curves after initial vetting removed all detected significant signals. This lack of signals in the FFI light curves is likely due to imperfect background subtraction or correction, the sparser sampling of the longer cadence, or instrumental effects that would otherwise be removed by the SPOC pipeline for 2 minute cadence data. It is worth noting that there were some marginal (5.0 $<$ SDE $<$ 7.0) detections of periodic signals in both cadences. Due to the lack of signals recovered in the FFI data, subsequent data, methods, and results will pertain to 2-minute cadence data only. All signals with a SDE $>$ 7.0 were passed along for further vetting and validation as described in the following sections.

To ensure that the search did not miss signals that may have been inadvertently diluted by the flattening procedure, we also searched the PDC and \texttt{eleanor}-corrected light curves without flattening as well as the Simple Aperture Photometry (SAP), \texttt{eleanor} PSF, and \texttt{eleanor} PCA light curves, with and without flattening. We found 2,434 non-HJ significant signals, 43 of which could not be immediately thrown out as noise. Nine of the signals that could not be ruled out as noise were also found in the search of the flattened PDC light curves. The remaining 34 signals were heavily scrutinized prior to any further analysis and all of them were rejected as potential candidates on the basis of transits overlapping with unsubtracted in-transit points from the HJs, unrealistic planet parameters, or $>$50$\%$ of transit events on the edge of the observation window or in observation gaps.

Nine of the 126 HJs with 2 minute cadence data were not recovered by the TLS search with a SDE $>$ 7.0. Of these 9 HJs that were not recovered, 7 HJs were around host stars with $Tmag$ $\gtrsim$ 13.1, 1 HJ was located near the galactic disk with a high degree of crowding, and 1 HJ orbits a host star with extreme stellar variability. Six of the 58 HJs with only 30 minute FFI cadence were not recovered. Four of these HJs were around host stars with $Tmag$ $\gtrsim$ 13.8, 1 HJ was around a host star with large stellar variability, and 1 HJ was located near the galactic plane and suffered from a high degree of crowding.

\subsection{\texttt{DAVE} Analysis} \label{ssec:dave}

The \texttt{DAVE} tool is an open-source Python package that wraps many common exoplanet vetting tools (\citealp{coughlin14}, e.g. \textit{Robovetter}) into one streamlined pipeline. This software has been extensively used to vet planet candidates from the \textit{Kepler} mission \citep{hedges19, kostov2019a, kostov2019b} and the TESS mission \citep{crossfield19, kostov2019b}. \texttt{DAVE} performs light-curve based vetting tests (odd-even transit comparison, a search for transit-like features due to light curve modulations, secondary eclipse checks) and image-based vetting tests (photocenter shift during transit).

Each significant periodic signal recovered with TLS was passed through \texttt{DAVE} and those that failed any of its modules were flagged for further inspection and removed from the analysis pipeline. If \texttt{DAVE} analysis flagged a signal in error, the signal was returned to the general pool of vetted signals. In total, 50 out of the initial 242 recovered signals passed DAVE vetting, although not all DAVE modules were able to run successfully for each signal. This is because in many cases, the transits of potential new signals overlapped with or were too close to those of the HJs, causing \texttt{DAVE} to run each module for the HJ multiple times instead of once for each of the signals in the light curve. This issue mostly affected the light-curve vetting tests and the image-based centroid vetting tests ran successfully for the majority of the target systems.

\subsection{VESPA Validation} \label{ssec:vespa}

To complement \texttt{DAVE} analysis, we used \texttt{vespa} \citep{2012ApJ...761....6M, 2015ascl.soft03011M} to calculate the false-positive probabilities of each signal. When provided stellar parameters, celestial coordinates, and orbital parameters, this package compares transit-like signals to a variety of astrophysical false-positive scenarios including an unblended eclipsing binary (EB), a blended background EB, a hierarchical companion EB, and the 'double-period' scenarios for each of these EB possibilities. All stellar parameters used in the \texttt{vespa} analysis were queried from the TIC for each individual target system \citep{stassun2019revised}. Orbital and planetary parameters from the TLS search output were used as inputs for the first round of \texttt{vespa} validation for each signal. If signals were successfully validated by \texttt{vespa} and proceeded to the \texttt{exoplanet} modeling step outlined in Figure \ref{fig:pipeline_flowchart}, the orbital and planetary parameters from this modeling were used for a secondary round of \texttt{vespa} validation. The light curves used in the TLS search outlined in Section \ref{ssec:signal_search} were folded according to the best-fit orbital period and mid-transit time (t0). These phase-folded light curves were used in the \texttt{vespa} analysis, oftentimes binned to reduce the scatter in the light curve and prevent an invalid fit of the transit shape or unreasonable posterior values.
 
A \texttt{vespa} input parameter of particular sensitivity is the maximum aperture radius (\textit{maxrad}) interior to which the signal must originate. This parameter strongly affects the likelihoods of the background eclipsing binary scenarios that \texttt{vespa} considers and is very dependent on sky position and the instrument used. We queried the \textit{Gaia} DR2 catalog within the SPOC pipeline extraction aperture to identify nearby background sources that are within 6 magnitudes of the target in the $G_{RP}$ band (630 - 1050 nm) \citep{brown2018gaia}. This band was chosen for its large overlap with the TESS band (600 - 1000 nm) \citep{vanderspek2018tess}. The \textit{maxrad} parameter was then set to the outermost background source meeting this criterion within the extraction aperture. If no background sources within the extraction aperture met this criterion, the \textit{maxrad} parameter was conservatively set to 2" to account for possible target position offset and the resolution of \textit{Gaia}. We note that, given the high resolution of \textit{Gaia}, this \textit{maxrad} parameter could be reduced to an even lower value in more thorough treatments of individual sources.
 
The maximum depth for a possible secondary eclipse (\texttt{vespa}'s \textit{secthresh} parameter) was used from the secondary eclipse depth output from \texttt{DAVE} if \textit{modshift} successfully ran a given target. Otherwise, a quick TLS search was performed for secondary transit-like features and that depth was used for the \textit{secthresh} parameter.

This \texttt{vespa} routine was repeated up to 25 times for each signal binned with values between 1 and 25 data points per bin. This was done to mitigate any variations between individual \texttt{vespa} simulations and because oftentimes \texttt{vespa} was unable to correctly fit a transit shape to the light curve or there was an error with calculating the posterior distributions of one or more astrophysical scenarios. The \texttt{vespa} simulation with the lowest binning value that did not produce an error was kept for each signal.

If the false positive probability (FPP) from the kept \texttt{vespa} simulation $<$1$\%$, we considered the signal to be statistically validated and the transit signal classified as a bona fide planet candidate. These FPP values calculated are likely upper limits since \texttt{vespa} analysis does not account for any likelihood increase due to a multiplicity boost from the confirmed HJ in each system. A multiplicity boost is the decrease in the FPP that a planet candidate gets from having other confirmed planets in the system since statistical work has demonstrated that systems containing multiple transit signals are more likely to be true planets than systems with only a single periodic signal \citep{Lissauer2012}. The exact multiplicity boost has not yet been calculated for TESS, therefore we elected to keep signals with a FPP value $<$10$\%$ in the analysis as possible ``marginal" signals in case any of them could pass below the 1$\%$ threshold when the multiplicity boost is calculated\footnote{We note that the multiplicity boost for TESS planets in general may be different from that of HJ systems specifically since there is strong evidence that HJ systems exhibit different planet clustering behavior than other planetary systems. Such a calculation is outside the scope of this work and marginal signals are included in this study to be as thorough as possible in light of an unknown multiplicity boost.}. This could very well be the case if the TESS multiplicity boost is at all similar to that calculated for \textit{Kepler} in \cite{Lissauer2012}.

Of the 50 signals that passed DAVE vetting, 14 produced an FPP value of $<$10$\%$ for at least one of its \texttt{vespa} iterations with 3 of these 14 signals producing an FPP value $<$1$\%$. These 14 signals were passed to \texttt{exoplanet} for more detailed modeling.

It is worth noting that \texttt{vespa} only tests against the six astrophysical false positive scenarios and does not take into account potential contamination from instrumental effects. While the instrumental false alarm rate for TESS has yet to be calculated, the TESS detectors exhibit fewer electronic noise artifacts than \textit{Kepler}\footnote{\url{https://archive.stsci.edu/kepler/manuals/KSCI-19033-001.pdf} and \url{https://archive.stsci.edu/files/live/sites/mast/files/home/missions-and-data/active-missions/tess/_documents/TESS_Instrument_Handbook_v0.1.pdf}}, which this software was developed on and where the FPP $<$ 1$\%$ validation threshold was established \citep{coughlin14, Krishnamurthy2019, vanderspek2018tess}. Therefore, we believe it is safe to assume that the 1$\%$ FPP threshold still holds here for TESS. However, as we discuss in Section \ref{sec:exoplanet}, the 14 signals that we identified as passing the \texttt{vespa} validation were subsequently determined to be instrumental effects after detailed light curve modeling and further manual inspection. Since \texttt{vespa} only tests for astrophysical false positives, these instrumental effects would not necessarily have been caught by \texttt{vespa} as non-planetary signals.

\section{Determination of Precise Planet Parameters} \label{sec:exoplanet}

TLS uses a period and transit duration search grid calculated upon initialization based on stellar properties and light curve length that is used to find periodic transit-like signals in a light curve \citep{hippke2019transit}. This grid can be oversampled for greater precision in period and duration of a transit-like feature, however this can quickly become computationally expensive and may still not produce the most precise orbital parameters. To remedy this, we used the software \texttt{exoplanet} \citep{exoplanet:exoplanet} on only the periodic signals that passed through \texttt{DAVE} and \texttt{vespa}. \texttt{exoplanet} is a toolkit for probabilistic modeling of transit and radial velocity observations of exoplanets using PyMC3. This is a powerful and flexible program that can be used to build high-performance transit models and then sample them through Markov Chain Monte Carlo (MCMC) simulations to provide precise transit and orbital parameters.

\begin{figure*}
    \centering
    \includegraphics[width=1.0\textwidth]{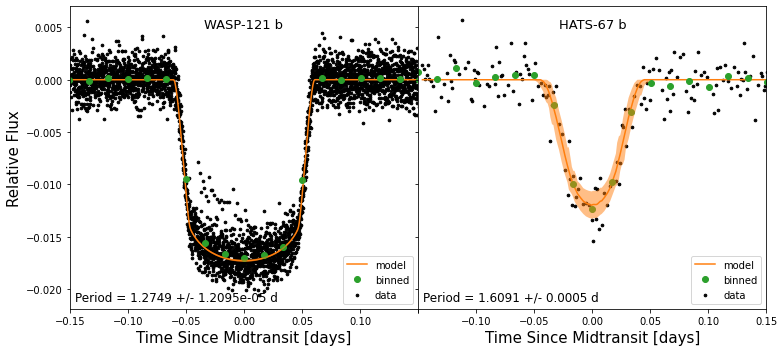}
    \caption{Examples of best-fit models with \texttt{exoplanet} for 2 minute cadence data (left, WASP-121 b) and 30 minute cadence data (right, HATS-67 b). Green points correspond to binning such that 19 points appear within the bounds of each axis. The transit model based on best-fit sampled posterior values is plotted in orange with light orange shading to represent the extent of 1$\sigma$ errors.}
    \label{fig:exoplanet_examples}
\end{figure*}

Examples of \texttt{exoplanet}-sampled HJ transits from our analysis is shown in Figure \ref{fig:exoplanet_examples} with both 2 minute and 30 minute cadence data. \texttt{exoplanet} was run to determine planet parameters for each of the 169 recovered HJs and the 14 new signals statistically validated to a FPP $<$10$\%$ by \texttt{vespa}. Through these detailed light curve model fits, we concluded that none of these signals arise from planets. Instead, they are noise, systematics, or integer multiples of improperly-subtracted HJ transits based on the best-fit transit parameters. These signals likely passed through the initial light curve detrending since they were variable on the same timescale as a typical transit duration and were sector- or CCD-specific features in the TESS data that were missed by the SPOC pipeline's detrending and our subsequent detrending with \texttt{lightkurve}. Furthermore, TLS may not have accurately determined the duration of the HJ transits in the system, causing the wings of the transit to be left behind after HJ transit subtraction.

Although no new promising planet candidates were found, we use the \texttt{exoplanet} models for these systems to provide a uniform set of updated orbital and planet parameters derived from TESS data for each of the HJs. These values can be found in Appendix \ref{app:orb_params}.

\begin{table}
    \caption{Table listing the number of new signals that passed each stage of the pipeline.}
    \label{tab:pipe_signals}
    \begin{tabular}{c|c}
    \hline
    \hline
        Pipeline Stage & $\#$ of Signals Passed \\
    \hline
        TLS Search & 242 \\
        \texttt{DAVE} Vetting & 50 \\
        \texttt{vespa} Validation & 14 \\
        \texttt{exoplanet} Modeling & 0 \\
    \end{tabular}
\end{table}

\section{Companion Rate Estimation} \label{sec:injection}

Although no new planet candidates were discovered through our search, it is still possible to place an upper limit on the rate of companion planets per HJ in the sample of systems used here. To do this, we need to know the efficiency at which our pipeline can recover transit signals so that we can correct our non-detection of additional companion planets for completeness of the search. In order to determine this efficiency, we performed a series of transit injections into light curves with known HJs that were then run through our implementation of TLS to probe the recovery rate of this method within different parameter spaces. For our detection efficiency calculation and subsequent estimation of the rate of companion planets per HJ, we only consider the 168 HJs that we are able to recover with our search pipeline after removing TOI-1130 since it was added after the target list was generated and would bias the statistical analysis. Furthermore, transit injection and successful recovery serves as an independent check and validation of our transit search algorithm implementation.

\begin{figure*}[!ht]
    \centering
    \includegraphics[width=0.99\textwidth]{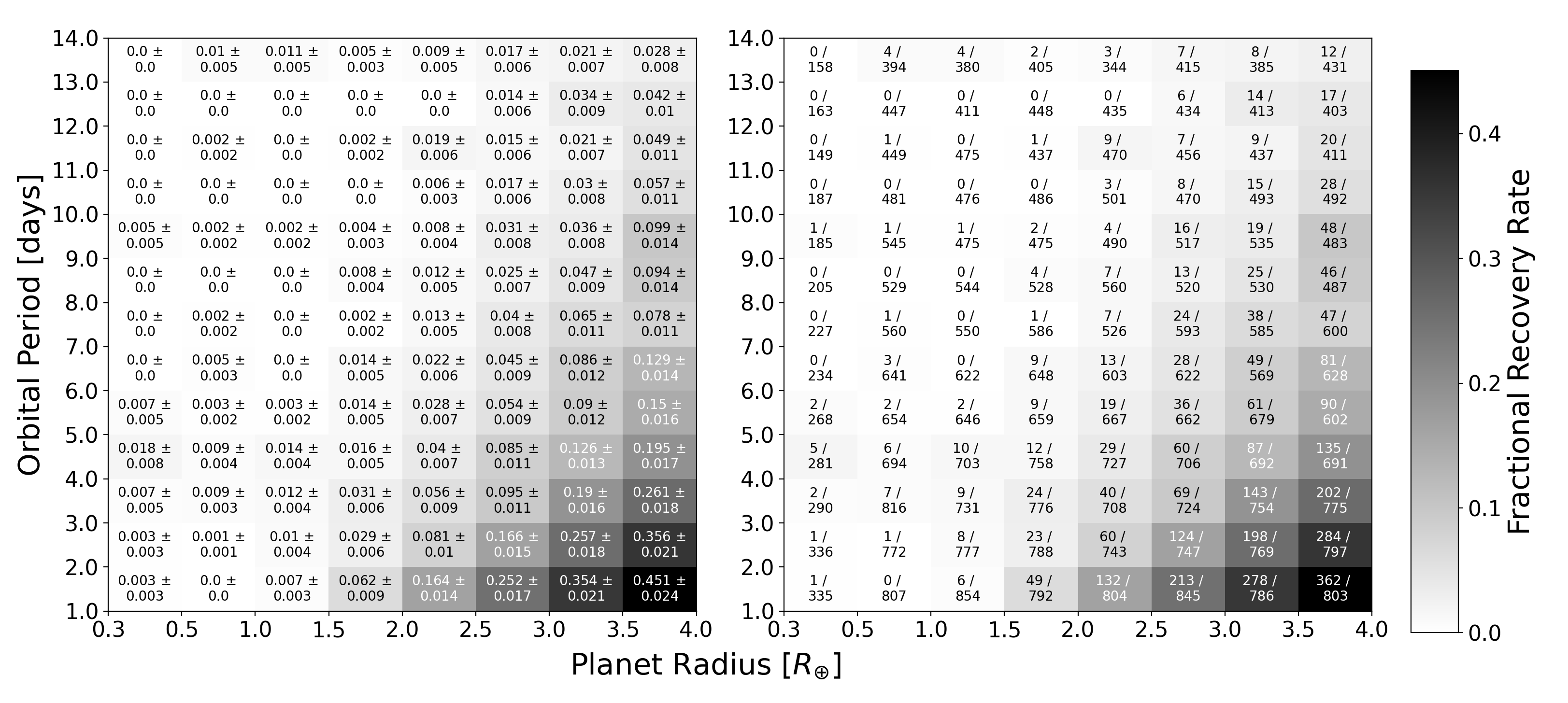}
    \caption{Grids displaying the recovery rates for each bin in the entire period and radius space simulated. The color in each cell denotes the fraction of injected planets recovered, averaged over all host star magnitudes and simulated inclinations in that slice of period-radius parameter space. Values of $<$5e-4 were rounded to 0 for clarity. Simulated transiting planets of all S/N are included. See Section \ref{sec:injection} for further details. \textit{Left:} Values in the cells denote the recovery rates of that slice of parameter space. Errors represent the square root of the number of successfully recovered injections divided by the total number of injections in each cell. \textit{Right:} Values in the cells denote the number of recovered injections over the total number of injections in each slice of parameter space.}
    \label{fig:recrates_allsims}
\end{figure*}

\begin{figure*}
    \centering
    \includegraphics[width=0.8\textwidth]{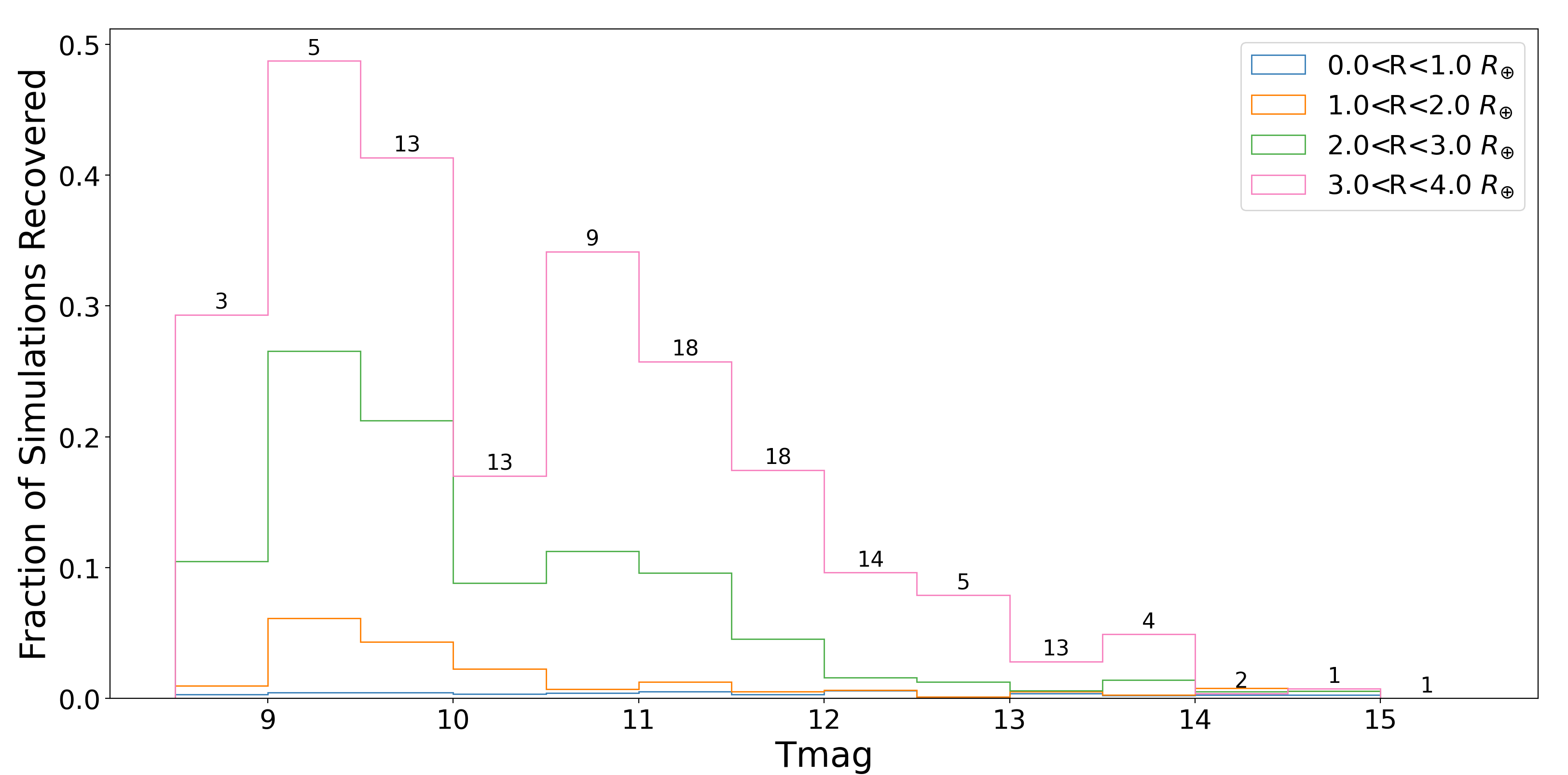}
    \caption{A histogram illustrating the recovery rates of simulated planets of different radii as it depends on the $Tmag$ of the host star. Each line represents a different slice of the planetary radii simulated, binned into corresponding bins in magnitude space of width $Tmag$=0.5. The histogram values are averaged over all inclinations and periods and contain signals of all S/N. The numbers above the highest histogram value in each bin represent the number of HJ light curves that were injected into in that $Tmag$ bin. This is included to illustrate how variability, artifacts, or quirks of individual light curves in $Tmag$ bins with few HJ light curves can affect the overall recovery rates. These numbers are also included to highlight the aim of these simulations to probe the sensitivity of this search in this particular data set rather than HJ light curves as a whole. The geometric transit probability is not applied here in order to more clearly illustrate the effect of $Tmag$ and individual stars on the detection efficiency. See Section \ref{sec:injection} for further explanation.}
    \label{fig:recrates_radhist}
\end{figure*}

To perform these simulations, we used the \texttt{Batman} Python package \citep{kreidberg2015batman} to generate artificial transits that were injected directly into the TESS PDC light curves of all 117 HJs that were recovered in the 2 minute cadence TESS data. The 9 PDC light curves in which the HJ was not recovered were not included in these simulations. By injecting simulated transits into real TESS data, we were able to obtain more realistic transit recovery scenarios than if synthetic light curves were used. 

We simulated $\sim$57,000 planet transits with randomly and uniformly sampled orbital period, planet radius, orbital inclination ($i$), injected into one of the 117 HJ light curves, randomly selected for each iteration. These HJ light curves orbit around host stars with 7.3 $\leq$ TESS magnitudes ($Tmag$) $\leq$ 17.8. The quadratic limb-darkening parameters and semi-major axis for each simulated planet were derived from the stellar parameters of the host star of the light curve that each transit was injected into. The orbital period was sampled between 1 day and 14 days to ensure that there were at least 2 transits in each $\sim$27-day TESS sector. The planetary radius was sampled between 0.3 $R_{\oplus}$ and 4 $R_{\oplus}$, the radius of one of the smallest exoplanets discovered \citep{barclay2013sub} and slightly larger than the largest HJ companion discovered (TOI-1130 b), respectively. The inclination was sampled from within the 1$\sigma$ uncertainties of the host HJ's inclination so that the planets could be considered coplanar. A circular orbit was assumed. Each of the simulations was run through our implementation of TLS using the same procedure outlined in \ref{ssec:signal_search} and compared with the simulated parameters. Any simulation where the strongest non-HJ signal with a SDE $>$7.0 matched the simulated period within its 1$\sigma$ errors was considered ``recovered." 

It is important to note that only combinations of parameters that produced transits of nonzero depth were considered. All of the simulations that were included in further analysis exhibited a transit of nonzero depth and none of the recovery rates include non-transiting cases.

Figure \ref{fig:recrates_allsims} displays the recovery rates for each segment of the orbital period and planet radius parameter space simulated across all inclinations and host star Tmag values. The highest recovery rate of $\sim$45.1$\%$ corresponds to the largest planets that transit the most frequently, with that of Earth-sized planets reaching no higher than 1.4$\%$. This sensitivity grid, in essence, represents the total fraction of planets within the $R_{p}$ and $P$ space that our pipeline was able to recover, regardless of signal strength. 

To quantify the dependence of the recovery rate on $P$ and $R_{p}$, we fit a double power law of the form $k P^{-\alpha} R_{p}^{-\beta}$ where k is a constant, and $\alpha$ and $\beta$ are the power law indices of $P$ and $R_{p}$. Using the \texttt{scipy.optimize} package to fit the function to the recovery rate grid, we found best fit values of $\alpha$=0.88 $\pm$ 0.03 and $\beta$=-2.32 $\pm$ 0.12, or a $P^{-0.88} R_{p}^{2.32}$ dependence for the recovery rates presented in Figure \ref{fig:recrates_allsims}.

The recovery rates are not uniform within each $R_{p}$ and $P$ cell as there is some dependence on parameters other than $R_{p}$ and $P$. TESS magnitude of the host star, in particular, strongly affects the recovery rate of the simulated planet transits. Figure \ref{fig:recrates_radhist} is included to illustrate how the recovery rates of various planet radii ranges vary in the host star TESS magnitude parameter space. Generally, brighter host stars correspond to higher recovery rates, although some TESS magnitude bins do not exhibit this due to the artifacts or mild stellar variability of individual light curves included within them that initial light curve flattening does not remove. Figure \ref{fig:recrates_radhist} includes the number of HJ light curves contained in each Tmag bin to illustrate how individual effects influenced Tmag bins with fewer light curves.

Comparing these recovery rates to that of the original TLS validation paper \citep{2019A&A...623A..39H}, we observe lower recovery rates. However, this is likely due to a combination of factors. TLS was originally designed and implemented for use with \textit{Kepler} data, which typically have greater photometric precision and longer time baselines than TESS data \citep{borucki2010kepler, ricker2015transiting}, arguably making it easier to recover small planets in \textit{Kepler} data since these factors cause small signals to have higher S/N than in TESS data. Additionally, TLS was validated on 1 $R_{\oplus}$ planet signals injected into artificial light curves with purely Gaussian noise and long baselines of 3 years. Therefore, it is logical that the shorter observation baseline of each TESS sector combined with non-Gaussian noise terms and lower photometric precision would produce recovery rates lower than the 93$\%$ stated by \cite{2019A&A...623A..39H}. Furthermore, as described in \cite{kruse2019detection}, mutual recovery rates of planet candidates of $\sim$60$\%$ are not uncommon from survey to survey. This again highlights the importance of multiple, independent searches of the same data set using separate methods so as to maximize the number of planet discoveries.

In order to estimate the rate of nearby companions to HJs, we addressed the problem through a Bayesian binomial framework following the methodology described in Appendix A of \cite{huang2016warm}. The likelihood of observing $N_{obs}$ companions to HJs from $N_{tot}$ systems with a multiplicity rate of $r_{m}$ can be expressed as B($N_{obs}|r_{m}$, $N_{tot}$). In our case, $N_{obs}$ is observable, $r_{m}$ is constrained given the data, and $N_{tot}$ is the number of HJs recovered in our sample ($N_{HJ}$) multiplied by a detection efficiency averaged across the entire parameter space ($\langle d_{eff} \rangle$) to correct for possible missed transiting planets according to Equation 5 in \cite{zhu2021exoplanet}. We can constrain $r_{m}$ by sampling the posterior space, given by the likelihood described above multiplied by some prior function for $r_{m}$. In this case, we adopt a uniform prior between 0 and 1 for the rate of nearby companions to HJs ($r_{m}$).

We perform a MCMC simulation using emcee \citep{2013PASP..125..306F} to sample this posterior space and constrain the rate of HJ multiplicity. For our MCMC model, we assumed that the companions transit and are coplanar with the HJ in the system (within the uncertainties of the HJ's inclination). For potential companions in the parameter space sampled here of 0.3$\leq$$R_{p}$$\leq$4 \rearth and 1$\leq$$P$$\leq$14 days, we calculate a $\langle d_{eff} \rangle$ of 4.8$\%$ and we obtain a HJ multiplicity rate of $7.3\substack{+15.2 \\ -7.3}\%$. If we shrink this potential companion parameter space to exclude companions of $R_{p}<$2.0 \rearth as in \cite{huang2016warm}, we obtain a HJ multiplicity rate of $4.2\substack{+9.1 \\ -4.2}\%$. These values represent the median of the distribution and the 90$\%$ confidence interval. Table \ref{tab:frequency} summarizes the rate of nearby companions for these two slices of the companion parameter space at various confidence intervals.

The values reported here are consistent with the values reported by both \cite{huang2016warm} and \cite{zhu2021exoplanet}. Both of these studies utilized the \textit{Kepler} sample in their estimation of companion rate, which has a higher detection efficiency due to its greater photometric precision, but a smaller HJ sample size than the TESS HJ sample used in this study.

\begin{table}
    \caption{The rate of companions per HJ at various confidence intervals for both the full range of potential companions with 0.3$\leq$$R_{p}$$\leq$4 \rearth and the narrower range 2.0$\leq$$R_{p}$$\leq$4 \rearth.}
    \label{tab:frequency}
    \begin{tabular}{c|c|c}
    \hline
    \hline
        Percentile & 0.3$\leq$$R_{p}$$\leq$4 \rearth & 2.0$\leq$$R_{p}$$\leq$4 \rearth \\
    \hline
        50$\%$ & 7.2$\%$ & 4.2$\%$ \\
        68$\%$ & 11.8$\%$ & 6.8$\%$ \\
        90$\%$ & 22.0$\%$ & 13.3$\%$ \\
        95$\%$ & 27.8$\%$ & 17.1$\%$ \\
        99$\%$ & 39.4$\%$ & 25.0$\%$ \\
    \end{tabular}
\end{table}

We note that the TOI-1130 system was not included in this part of the analysis. Including it would bias the result since the system was only added into the search sample after the original list had been generated on the basis that TOI-1130 is contained within the TESS field of view and exhibited the type of system structure that this study was searching for. For comparison's sake, the same statistical estimation of HJ companion rate was performed including TOI-1130, resulting in a rate of $17.7\substack{+18.6 \\ -11.7}\%$ across the whole parameter space probed and a rate of $10.3\substack{+12.0 \\ -6.9}\%$   when excluding companions with $R_{p}<$2.0 \rearth. The upper and lower limits on these values represent the 90$\%$ and 10$\%$ confidence intervals, respectively.

\section{Discussion} \label{sec:discussion}

\subsection{Comparison to Other Searches} \label{ssec:pipeline_comparison}
Both the SPOC pipeline \citep{Jenkins2016} and the MIT Quicklook Pipeline (QLP) \citep{huang2020photometry, huang2020photometry2} searched each of the HJ systems included in this study with their independent pipelines and also returned no new planet candidates in systems with confirmed HJs. Furthermore, smaller scale studies on subsets of the HJ population - such as that of \cite{steffen2012kepler} and \cite{2020arXiv201011977M} -  also find no nearby companions and sometimes complement the photometric analysis with radial velocity data. Our independent search of the HJ population in the southern ecliptic hemisphere provides compelling support for a lack of planetary companions nearby these HJs down to the photometric precision of TESS. Each pipeline searched using different search algorithms, ranges of orbital period, and light curve processing, thus maximizing the parameter space within which new signals can be detected in these target systems.

The QLP did discover an entirely new HJ system in the Cycle 1 TESS FFIs. This system - TOI-1130 - does indeed contain a HJ along with an inner companion \citep{huang2020tess}. This system was not originally included in our study since it was not confirmed prior to the initial target list but was included in subsequent companion searches in this study. However, TOI-1130 was not included in our statistical analysis of the HJ population since it was not discovered prior to the start of this study and its addition would bias the results. The pipeline presented here was able to recover both the HJ signal and the companion planet signal of this new TOI-1130 without need for any additional systematics or noise correction, despite the fact that no new significant signals were recovered from FFI data by our pipeline otherwise. 

The agreement of the results from the QLP and SPOC pipeline with that presented here, although not necessarily expected, serves as an excellent check of our pipeline and the validity of the TOI-1130 system, especially since the other two systems harboring HJs with nearby companions (WASP-47 and Kepler-730) were not contained within the Cycle 1 TESS data. Additionally, since there has as of yet been no direct comparison between the sensitivity of TLS and the SPOC search pipeline, these results can serve as an indication that the sensitivities of this TLS search, the SPOC pipeline, and the QLP are comparable. This is of particular interest given the results from the validation of TLS suggesting that TLS has a $\sim$17$\%$ higher detection efficiency than BLS, which is used by the QLP \citep{2019A&A...623A..39H}.

Although this study probes a slightly different parameter space of nearby companions to hot Jupiters than \cite{huang2016warm}, the calculated values for the rate of companions to HJs are consistent with \cite{huang2016warm} who reported $1.1\substack{+13.3 \\ -1.1}\%$ compared to $7.3\substack{+15.2 \\ -7.3}\%$ calculated by this study. The larger period and radius space probed by this study more closely matches the parameter space studied by \cite{zhu2021exoplanet}, which reports a value of $\sim$2$\%$ with a 95$\%$ confidence interval of 9.7$\%$ for the rate of companions nearby HJs. This is also consistent with our result. Both of these previous studies utilized the \textit{Kepler} sample of HJs in their determination of the rate of companions nearby HJs, which has greater photometric precision - and therefore higher detection efficiencies - but a smaller sample size of HJs. Our study marks the first uniform calculation of this rate with the TESS HJ sample.

\subsection{Implications for HJ Formation} \label{ssec:formation_implications}
The lack of any additional new validated planets in the HJ systems we searched supports previous results that indicate a general lack of planets in nearby orbits to HJs \citep{steffen2012kepler}. The only exceptions to this trend are WASP-47 d, WASP-47 e, Kepler-730 c, and the recently-discovered TOI-1130 b \citep{becker2015wasp, canas2019kepler, huang2020tess}. This lack of nearby planets to HJs is in stark contrast to the ``warm Jupiters" (WJs), a class of planet similar to HJs, but with an orbital period between 10 and 200 days. Approximately 50$\%$ of WJs have nearby companion planets compared to the $1.1\substack{+13.3 \\ -1.1}\%$ of HJs with companions as reported by \cite{huang2016warm} or the rate of $7.3\substack{+15.2 \\ -7.3}\%$ reported here, despite the orbital period cutoff distinguishing the two classes being somewhat arbitrary. In fact, as discussed in \cite{huang2020tess}, the period distribution of giant planets with nearby companions appears continuous from the shortest period of the three aforementioned HJs (WASP-47 b) through the WJ periods. This could suggest that these handful of HJs with companions formed in a similar manner to the slightly cooler WJs while the rest of the HJs formed via a separate pathway.

Some formation scenarios make specific predictions for the occurrence of nearby companion planets. In the case of formation through high eccentricity migration (HEM) where the giant planet arrives at its current position via gravitational scattering of other bodies in the system, the likelihood that nearby planets exist is low due to the disruptive nature of the planet migration \citep{mustill2015destruction}. In the case of a disk migration where the entire protoplanetary disk migrates inward, companion planets would be more likely to survive but would also likely exhibit orbital resonances with one another \citep{lee2002dynamics, raymond2006exotic}.

However, while a lack of nearby companion planets cannot definitively determine the pathway through which each system formed, this characteristic may aid in classifying portions of the HJ population when combined with additional evidence. Some formation scenarios that challenge current HJ formation theories could benefit from knowing that HJs are lonely with a greater amount of certainty. For instance, there are possible situations where companion planets are retained despite HEM \citep{fogg2007formation} or situations where HEM cannot explain the observed dynamics in a handful of known HJs \citep{dawson2014photoeccentric}. Constraining the presence of nearby companions to HJs may help in determining the dynamical histories of some of these scenarios that complicate our theories on HJ formation.

Although these systems are proving to be quite rare, it is important to continue to search for HJ systems with closely-orbiting companions so that comparisons can be drawn between this unique subset of systems and the wider sample of HJs/WJs. Additional discoveries of HJ systems with nearby companions would contribute to a better understanding of how these systems formed and if the mechanism differs from other portions of the HJ/WJ population. Furthermore, a scaled-up statistical analysis including the larger TESS field and all three systems with known nearby companions to HJs would provide a companion rate per HJ that is much more representative of the HJ population as a whole since the rate reported here only considers the HJs of the TESS southern ecliptic hemisphere in TESS Cycle 1.

\section{Summary} \label{sec:conclusion}

In this paper, we present the results of an independent, uniform search for companions to HJs in TESS southern ecliptic hemisphere data (Sectors 1-13). Our investigation and results are summarized here:

\begin{itemize}
    \item We searched the TESS light curves of 184 systems with HJs of $R_{p}>$ 8\rearth using Transit Least Squares with both the default and grazing transit shapes.
    
    \item New signals recovered by the Transit Least Squares search with a signal detection efficiency $>$ 7.0 were passed through \texttt{DAVE} and \texttt{vespa} for vetting and validation.
    
    \item There were zero new signals with $P <$ 14 days statistically validated as planet candidates to a false positive probability $<$ 1$\%$ in either the 2-minute cadence SPOC PDC light curves or the 30-minute TESS FFI light curves. We cannot rule out the existence of transiting companions with $P >$ 14 days, however.
    
    \item We probed the detectable parameter space of potential small planet signals using our pipeline, finding a dependence of recovery rates proportional to $R_{p}^{2.32}P^{-0.88}$. We found a strong dependence on magnitude and activity of the host star as well for all recovery rates.
    
    \item We performed a statistical analysis to estimate a rate of $7.3\substack{+15.2 \\ -7.3}\%$ planets within 0.3$\leq$$R_{p}$$\leq$4 \rearth and 1$\leq$$P$$\leq$14 days per HJ.
    
    \item A lack of new companion planets to HJs down to the photometric precision of TESS provides further evidence for the ``loneliness" of HJs out to $P$=14 days and HEM as a plausible formation mechanism for a large portion of the HJ population. This is in contrast to warm Jupiters where nearby companions are common, suggesting possible different formation mechanisms for the two populations.
    
    \item These search results suggest that the sensitivities of the SPOC search pipeline, MIT's QLP, and this TLS pipeline are comparable in the search for small companions to HJs.
\end{itemize}

This work constitutes a first step in comprehensively searching every HJ observed by TESS. Similar studies of the HJ systems observed by TESS in its survey of the northern ecliptic hemisphere will be beneficial for further exploring potential HJ formation mechanisms. Furthermore, TESS has recently started its extended mission where it is effectively repeating its survey of the southern and northern ecliptic hemispheres and will also survey part of the ecliptic plane for the first time.

As an additional component of this work, we provide updated transit ephemerides for each HJ with TESS 2-minute Cycle 1 data in Appendix \ref{app:orb_params}. The majority of both the orbital period and the planetary radius value agree within 1$\sigma$ errors with published values. For a subset of these HJs, the planetary radius is better constrained with smaller uncertainties than published values. The eccentricity values are calculated based on stellar density according to the prescription in \cite{dawson2012photoeccentric} and are generally slightly higher than in the literature. This parameter is calculated and not directly sampled, so is subject to larger uncertainty than a sampled parameter. These can aid in follow-up observations and studies of the HJs themselves, since in the absence of planetary companions, further study of the HJs in these systems becomes even more important in constraining their formation processes.

\acknowledgments

We thank D. Deming, E. Kempton, F. Adams, and the anonymous referee for their valuable feedback and comments on this study. 

This paper includes data collected by the TESS mission, which are publicly available from the Mikulski Archive for Space Telescopes (MAST) and produced by the Science Processing Operations Center (SPOC) at NASA Ames Research Center. This research effort made use of systematic error-corrected (PDC-SAP) photometry. Funding for the TESS mission is provided by NASA's Science Mission directorate. Resources supporting this work were provided by the NASA High-End Computing (HEC) Program through the NASA Advanced Supercomputing (NAS) Division at Ames Research Center for the production of the SPOC data products. 

This research has made use of the NASA Exoplanet Archive and Exoplanet Follow-up Observation Program website, which is operated by the California Institute of Technology, under contract with the National Aeronautics and Space Administration under the Exoplanet Exploration Program. This work has made use of data from the European Space Agency (ESA) mission {\it Gaia} (\url{https://www.cosmos.esa.int/gaia}), processed by the {\it Gaia} Data Processing and Analysis Consortium (DPAC, \url{https://www.cosmos.esa.int/web/gaia/dpac/consortium}). Funding for the DPAC has been provided by national institutions, in particular the institutions participating in the {\it Gaia} Multilateral Agreement. 

B.J.H., K.D.C., and V.K. acknowledge support from the GSFC Sellers Exoplanet Environments Collaboration (SEEC), which is funded in part by the NASA Planetary Science Division’s Internal Scientist Funding Model. B.J.H. also acknowledges support from the Future Investigators in NASA Earth and Space Science and Technology (FINESST) program - grant 80NSSC20K1551 - and support by NASA under award number 80GSFC21M0002.

This research made use of Lightkurve, a Python package for Kepler and TESS data analysis (Lightkurve Collaboration, 2018). This research made use of \textsf{exoplanet} \citep{exoplanet:exoplanet} and its
dependencies \citep{exoplanet:agol19, exoplanet:astropy13, exoplanet:astropy18,
exoplanet:exoplanet, exoplanet:foremanmackey17, exoplanet:foremanmackey18,
exoplanet:kipping13, exoplanet:kipping13b, exoplanet:luger18, exoplanet:pymc3,
exoplanet:theano}. 

\software{Astropy \citep{exoplanet:astropy18}, BATMAN \citep{kreidberg2015batman}, DAVE \citep{kostov2019a}, eleanor \citep{feinstein2019eleanor}, exoplanet \citep{exoplanet:exoplanet}, Lightkurve \citep{2018ascl.soft12013L}, Transit Least Squares \citep{2019A&A...623A..39H}, vespa \citep{2015ascl.soft03011M}}

\facilities{NASA Exoplanet Archive, TESS, MAST, \textit{Gaia}}

\bibliography{main}{}
\bibliographystyle{aasjournal}

\appendix
    
\section{Best-Fit Planet and Orbital Parameters for Full HJ Target List} \label{app:orb_params}

\begin{longrotatetable}
\begin{deluxetable}{cccccccccc}
\tabletypesize{\footnotesize}
\tablecaption{Best fit parameters from \texttt{exoplanet} simulations for all of the 169 detected HJs contained in our dataset. Errors quoted are 1$\sigma$ values. Targets that were not recovered by our pipeline (as discussed in Section \ref{ssec:signal_search}) are not included in this table. Transit depth values are calculated based off of the sampled $R_{p}/R_{*}$. Planet radii are calculated based off the sampled $R_{p}/R_{*}$ and the stellar radius and errors in the TIC. Eccentricity is calculated from stellar density based on the method outlined in \cite{dawson2012photoeccentric}. Some values for stellar radius and stellar density were taken from ExoFOP user-uploaded values if they were not available in the TIC\tablenotemark{4}. Values left blank were NaN due to missing or unconstrained values in the TIC and ExoFOP. This table is available in machine-readable format in the extended materials. \label{tab:ephemeris}}

\tablehead{ \colhead{TIC ID} & \colhead{Common Name} & \colhead{t0} & \colhead{Period} & \colhead{Transit Depth} & \colhead{Transit Duration} & \colhead{Radius} & \colhead{$R_{p}$/$R_{*}$} & \colhead{Impact Parameter} & \colhead{Eccentricity} \\
\colhead{} & \colhead{} & \colhead{(BJD + 2457000)} & \colhead{(days)} & \colhead{(ppm)} & \colhead{(days)} & \colhead{($R_{J}$)} & \colhead{} & \colhead{} & \colhead{} }

\startdata
1129033 & WASP-77 A b & 1410.98473 $\pm$ 9e-05 & 1.36003 $\pm$ 1e-05 & 14356 $\pm$ 175 & 0.0799 $\pm$ 0.00034 & 1.089 $\pm$ 0.05 & 0.11982 $\pm$ 0.00073 & 0.344 $\pm$ 0.04 & $0.16\substack{+0.32 \\ -0.12}$ \\
1528696 & NGTS-6 b & 1438.40435 $\pm$ 0.00103 & 0.88205 $\pm$ 6e-05 & 17826 $\pm$ 7380 & 0.023 $\pm$ 0.00524 & 1.689 $\pm$ 0.698 & 0.13351 $\pm$ 0.02764 & 0.823 $\pm$ 0.201 & $0.6\substack{+0.28 \\ -0.36}$ \\
4616072 & HATS-45 b & 1469.05829 $\pm$ 0.00172 & 4.18794 $\pm$ 0.00028 & 11519 $\pm$ 992 & 0.10869 $\pm$ 0.00347 & 1.292 $\pm$ 0.08 & 0.10733 $\pm$ 0.00462 & 0.383 $\pm$ 0.211 & $0.36\substack{+0.28 \\ -0.2}$ \\
6663331 & HATS-28 b & 1659.81120 $\pm$ 0.00149 & 3.18083 $\pm$ 0.00031 & 18887 $\pm$ 1068 & 0.07853 $\pm$ 0.00238 & 1.268 $\pm$ --- & 0.13743 $\pm$ 0.00389 & 0.275 $\pm$ 0.173 & --- \\
7088246 & HATS-23 b & 1656.50839 $\pm$ 0.00275 & 2.16027 $\pm$ 0.00041 & 19541 $\pm$ 8522 & 0.05194 $\pm$ 0.00944 & 1.428 $\pm$ --- & 0.13979 $\pm$ 0.03048 & 0.616 $\pm$ 0.298 & --- \\
12862099 & WASP-44 b & 1386.57835 $\pm$ 0.00069 & 2.42381 $\pm$ 0.00014 & 13794 $\pm$ 917 & 0.08205 $\pm$ 0.00184 & 1.102 $\pm$ 0.072 & 0.11745 $\pm$ 0.0039 & 0.386 $\pm$ 0.19 & $0.34\substack{+0.29 \\ -0.18}$ \\
13021029 & WASP-61 b & 1439.12368 $\pm$ 0.00088 & 3.85588 $\pm$ 0.00023 & 8940 $\pm$ 301 & 0.14984 $\pm$ 0.00193 & 1.318 $\pm$ 0.057 & 0.09455 $\pm$ 0.00159 & 0.274 $\pm$ 0.157 & $0.26\substack{+0.31 \\ -0.15}$ \\
13349647 & WASP-36 b & 1518.50523 $\pm$ 0.00042 & 1.53738 $\pm$ 5e-05 & 18584 $\pm$ 891 & 0.06391 $\pm$ 0.0017 & 1.273 $\pm$ 0.066 & 0.13632 $\pm$ 0.00327 & 0.67 $\pm$ 0.058 & $0.19\substack{+0.32 \\ -0.14}$ \\
14344979 & WASP-183 b & 1544.52324 $\pm$ 0.00209 & 4.11226 $\pm$ 0.00069 & 25779 $\pm$ 9932 & 0.06245 $\pm$ 0.01228 & 1.431 $\pm$ 0.284 & 0.16056 $\pm$ 0.03093 & 0.751 $\pm$ 0.23 & $0.24\substack{+0.34 \\ -0.18}$ \\
14614418 & HATS-18 b & 1571.2146 $\pm$ 0.00067 & 0.8378 $\pm$ 4e-05 & 21104 $\pm$ 798 & 0.06974 $\pm$ 0.00123 & 1.437 $\pm$ --- & 0.14527 $\pm$ 0.00275 & 0.216 $\pm$ 0.135 & --- \\
14661418 & HATS-22 b & 1575.71273 $\pm$ 0.0008 & 4.72312 $\pm$ 0.0003 & 19825 $\pm$ 1694 & 0.07885 $\pm$ 0.00242 & 1.087 $\pm$ 0.092 & 0.1408 $\pm$ 0.00601 & 0.439 $\pm$ 0.195 & $0.45\substack{+0.25 \\ -0.19}$ \\
15445551 & WASP-87 b & 1571.34985 $\pm$ 0.00029 & 1.68279 $\pm$ 2e-05 & 8013 $\pm$ 147 & 0.11181 $\pm$ 0.00068 & 1.366 $\pm$ 0.064 & 0.08952 $\pm$ 0.00082 & 0.576 $\pm$ 0.034 & $0.2\substack{+0.32 \\ -0.14}$ \\
16288184 & K2-237 b & 1628.98137 $\pm$ 0.00085 & 2.18077 $\pm$ 0.00014 & 15701 $\pm$ 676 & 0.1107 $\pm$ 0.00174 & 1.55 $\pm$ 0.084 & 0.1253 $\pm$ 0.0027 & 0.206 $\pm$ 0.13 & $0.29\substack{+0.31 \\ -0.16}$ \\
17746821 & HAT-P-50 b & 1493.16907 $\pm$ 0.00137 & 3.12183 $\pm$ 0.00031 & 6290 $\pm$ 463 & 0.13723 $\pm$ 0.00337 & 1.335 $\pm$ 0.096 & 0.07931 $\pm$ 0.00292 & 0.592 $\pm$ 0.182 & $0.27\substack{+0.32 \\ -0.19}$ \\
19684256 & HATS-52 b & 1518.45078 $\pm$ 0.00271 & 1.36673 $\pm$ 0.00026 & 16634 $\pm$ 2456 & 0.07719 $\pm$ 0.00565 & 1.282 $\pm$ --- & 0.12897 $\pm$ 0.00952 & 0.415 $\pm$ 0.24 & --- \\
22529346 & WASP-121 b & 1491.99869 $\pm$ 0.00013 & 1.27493 $\pm$ 1e-05 & 15413 $\pm$ 105 & 0.10713 $\pm$ 0.0003 & 1.839 $\pm$ 0.073 & 0.12415 $\pm$ 0.00042 & 0.073 $\pm$ 0.05 & $0.3\substack{+0.3 \\ -0.15}$ \\
29344935 & HATS-14 b & 1326.12717 $\pm$ 0.00123 & 2.76681 $\pm$ 0.00024 & 14053 $\pm$ 706 & 0.09477 $\pm$ 0.00226 & 1.046 $\pm$ 0.057 & 0.11854 $\pm$ 0.00298 & 0.345 $\pm$ 0.182 & $0.15\substack{+0.32 \\ -0.11}$ \\
29857954 & TOI-172 b & 1326.92282 $\pm$ 0.01872 & 9.47467 $\pm$ 0.0136 & 3198 $\pm$ 1135 & 0.19037 $\pm$ 0.03135 & 0.984 $\pm$ 0.182 & 0.05655 $\pm$ 0.01004 & 0.513 $\pm$ 0.288 & $0.43\substack{+0.26 \\ -0.27}$ \\
31858844 & WASP-123 b & 1658.55693 $\pm$ 0.00052 & 2.97755 $\pm$ 0.00012 & 40146 $\pm$ 824 & 0.11124 $\pm$ 0.00114 & 2.496 $\pm$ --- & 0.20036 $\pm$ 0.00206 & 0.093 $\pm$ 0.066 & --- \\
32499655 & HATS-44 b & 1440.25908 $\pm$ 0.00218 & 2.74395 $\pm$ 0.00019 & 15808 $\pm$ 5377 & 0.0457 $\pm$ 0.00616 & 0.963 $\pm$ 0.167 & 0.12573 $\pm$ 0.02138 & 0.638 $\pm$ 0.261 & $0.33\substack{+0.32 \\ -0.24}$ \\
32949762 & WASP-160 B b & 1468.98296 $\pm$ 0.00743 & 3.76779 $\pm$ 0.00229 & 10788 $\pm$ 2871 & 0.10583 $\pm$ 0.01125 & 0.884 $\pm$ 0.129 & 0.10387 $\pm$ 0.01382 & 0.466 $\pm$ 0.268 & $0.25\substack{+0.32 \\ -0.18}$ \\
33521996 & HATS-6 b & 1469.31156 $\pm$ 0.00064 & 3.3253 $\pm$ 0.00018 & 31174 $\pm$ 1665 & 0.07054 $\pm$ 0.0017 & 1.025 $\pm$ 0.041 & 0.17656 $\pm$ 0.00472 & 0.29 $\pm$ 0.16 & $0.3\substack{+0.3 \\ -0.12}$ \\
35516889 & WASP-19 b & 1544.41075 $\pm$ 0.00019 & 0.78885 $\pm$ 1e-05 & 22989 $\pm$ 660 & 0.05501 $\pm$ 0.0009 & 1.517 $\pm$ 0.075 & 0.15162 $\pm$ 0.00218 & 0.632 $\pm$ 0.028 & $0.3\substack{+0.3 \\ -0.14}$ \\
35857242 & WASP-138 b & 1413.31512 $\pm$ 0.00089 & 3.63468 $\pm$ 0.00022 & 7864 $\pm$ 321 & 0.15375 $\pm$ 0.00223 & 1.227 $\pm$ 0.065 & 0.08868 $\pm$ 0.00181 & 0.3 $\pm$ 0.17 & $0.2\substack{+0.32 \\ -0.14}$ \\
36352297 & CoRoT-1 b & 1469.06776 $\pm$ 0.00092 & 1.509 $\pm$ 0.00012 & 22482 $\pm$ 1148 & 0.08819 $\pm$ 0.00218 & 1.69 $\pm$ 0.129 & 0.14994 $\pm$ 0.00383 & 0.414 $\pm$ 0.174 & $0.2\substack{+0.33 \\ -0.14}$ \\
36440357 & CoRoT-4 b & 1475.33216 $\pm$ 0.00378 & 9.19807 $\pm$ 0.00519 & 16574 $\pm$ 1533 & 0.16468 $\pm$ 0.00717 & 1.406 $\pm$ --- & 0.12874 $\pm$ 0.00595 & 0.297 $\pm$ 0.191 & --- \\
36734222 & WASP-43 b & 1545.23052 $\pm$ 9e-05 & 0.81347 $\pm$ 1e-05 & 25925 $\pm$ 644 & 0.03998 $\pm$ 0.00056 & 1.187 $\pm$ 0.118 & 0.16101 $\pm$ 0.002 & 0.696 $\pm$ 0.014 & $0.41\substack{+0.29 \\ -0.19}$ \\
37168957 & CoRoT-18 b & 1470.17324 $\pm$ 0.00251 & 1.90026 $\pm$ 0.0004 & 28501 $\pm$ 2407 & 0.08525 $\pm$ 0.00376 & 1.352 $\pm$ 0.082 & 0.16882 $\pm$ 0.00713 & 0.279 $\pm$ 0.18 & $0.17\substack{+0.32 \\ -0.12}$ \\
37718056 & WASP-158 b & 1387.75207 $\pm$ 0.00733 & 3.65588 $\pm$ 0.00243 & 5967 $\pm$ 1515 & 0.14251 $\pm$ 0.01423 & 1.153 $\pm$ 0.158 & 0.07725 $\pm$ 0.0098 & 0.464 $\pm$ 0.267 & $0.31\substack{+0.3 \\ -0.2}$ \\
38846515 & WASP-100 b & 1571.79205 $\pm$ 0.00046 & 2.84933 $\pm$ 0.0001 & 7259 $\pm$ 128 & 0.14015 $\pm$ 0.001 & 1.471 $\pm$ 0.067 & 0.0852 $\pm$ 0.00075 & 0.62 $\pm$ 0.04 & $0.2\substack{+0.32 \\ -0.14}$ \\
42821097 & CoRoT-19 b & 1468.68564 $\pm$ 0.00392 & 3.89587 $\pm$ 0.00129 & 6873 $\pm$ 634 & 0.186 $\pm$ 0.00646 & 1.188 $\pm$ 0.087 & 0.08291 $\pm$ 0.00383 & 0.344 $\pm$ 0.212 & $0.22\substack{+0.32 \\ -0.16}$ \\
43647325 & WASP-35 b & 1440.12273 $\pm$ 0.0002 & 3.16157 $\pm$ 5e-05 & 14948 $\pm$ 291 & 0.11707 $\pm$ 0.0009 & 1.306 $\pm$ 0.064 & 0.12226 $\pm$ 0.00119 & 0.162 $\pm$ 0.098 & $0.23\substack{+0.32 \\ -0.15}$ \\
44745133 & HATS-61 b & 1415.7765 $\pm$ 0.01037 & 7.81693 $\pm$ 0.00689 & 3839 $\pm$ 1052 & 0.19581 $\pm$ 0.02083 & 0.986 $\pm$ 0.144 & 0.06196 $\pm$ 0.00849 & 0.496 $\pm$ 0.279 & $0.35\substack{+0.29 \\ -0.22}$ \\
46096489 & WASP-16 b & 1601.1914 $\pm$ 0.00041 & 3.11853 $\pm$ 9e-05 & 13137 $\pm$ 594 & 0.06599 $\pm$ 0.00228 & 1.194 $\pm$ 0.065 & 0.11462 $\pm$ 0.00259 & 0.808 $\pm$ 0.029 & $0.3\substack{+0.3 \\ -0.15}$ \\
47911178 & WASP-101 b & 1470.30411 $\pm$ 0.00029 & 3.58559 $\pm$ 9e-05 & 11937 $\pm$ 322 & 0.09354 $\pm$ 0.00138 & 1.431 $\pm$ 0.069 & 0.10925 $\pm$ 0.00147 & 0.719 $\pm$ 0.025 & $0.33\substack{+0.29 \\ -0.16}$ \\
50712784 & WASP-161 b & 1492.28167 $\pm$ 0.01508 & 5.40795 $\pm$ 0.00571 & 4944 $\pm$ 1642 & 0.204 $\pm$ 0.02819 & 1.091 $\pm$ 0.188 & 0.07031 $\pm$ 0.01168 & 0.51 $\pm$ 0.285 & $0.25\substack{+0.34 \\ -0.18}$ \\
52640302 & WASP-64 b & 1469.59033 $\pm$ 0.00047 & 1.57325 $\pm$ 3e-05 & 14725 $\pm$ 528 & 0.08851 $\pm$ 0.00125 & 1.293 $\pm$ 0.071 & 0.12135 $\pm$ 0.00218 & 0.238 $\pm$ 0.143 & $0.31\substack{+0.3 \\ -0.15}$ \\
52689469 & HATS-66 b & 1470.09153 $\pm$ 0.00495 & 3.14089 $\pm$ 0.00144 & 5853 $\pm$ 616 & 0.18869 $\pm$ 0.00862 & 1.291 $\pm$ --- & 0.07651 $\pm$ 0.00403 & 0.357 $\pm$ 0.221 & --- \\
53189332 & WASP-106 b & 1544.65032 $\pm$ 0.00137 & 9.29024 $\pm$ 0.00114 & 5690 $\pm$ 241 & 0.20615 $\pm$ 0.00297 & 1.039 $\pm$ 0.051 & 0.07543 $\pm$ 0.0016 & 0.35 $\pm$ 0.19 & $0.2\substack{+0.32 \\ -0.14}$ \\
53458803 & HATS-51 b & 1468.62066 $\pm$ 0.00419 & 3.34932 $\pm$ 0.00113 & 10306 $\pm$ 1478 & 0.12974 $\pm$ 0.00827 & 1.215 $\pm$ 0.106 & 0.10152 $\pm$ 0.00728 & 0.435 $\pm$ 0.244 & $0.25\substack{+0.31 \\ -0.17}$ \\
53735810 & WASP-66 b & 1548.25501 $\pm$ 0.00122 & 4.08664 $\pm$ 0.00047 & 6472 $\pm$ 235 & 0.17183 $\pm$ 0.00235 & 1.157 $\pm$ 0.062 & 0.08045 $\pm$ 0.00146 & 0.249 $\pm$ 0.156 & $0.17\substack{+0.33 \\ -0.12}$ \\
53750200 & HATS-41 b & 1468.98591 $\pm$ 0.0049 & 4.19633 $\pm$ 0.00139 & 4297 $\pm$ 859 & 0.07384 $\pm$ 0.00706 & 1.038 $\pm$ 0.114 & 0.06555 $\pm$ 0.00656 & 0.457 $\pm$ 0.266 & $0.73\substack{+0.13 \\ -0.27}$ \\
55092869 & KELT-11 b & 1549.07778 $\pm$ 0.00065 & 4.73605 $\pm$ 0.00025 & 2232 $\pm$ 62 & 0.27891 $\pm$ 0.00101 & 1.306 $\pm$ 0.054 & 0.04725 $\pm$ 0.00066 & 0.488 $\pm$ 0.074 & $0.23\substack{+0.32 \\ -0.16}$ \\
59843967 & HATS-4 b & 1468.52084 $\pm$ 0.00089 & 2.5168 $\pm$ 0.00019 & 15140 $\pm$ 686 & 0.09506 $\pm$ 0.00183 & 1.177 $\pm$ 0.071 & 0.12305 $\pm$ 0.00279 & 0.252 $\pm$ 0.152 & $0.31\substack{+0.3 \\ -0.16}$ \\
66818296 & WASP-17 b & 1630.86141 $\pm$ 0.00044 & 3.73544 $\pm$ 0.00015 & 15181 $\pm$ 390 & 0.15955 $\pm$ 0.00172 & 1.886 $\pm$ 0.109 & 0.12321 $\pm$ 0.00158 & 0.201 $\pm$ 0.118 & $0.28\substack{+0.31 \\ -0.17}$ \\
77031414 & WASP-173 A b & 1355.19582 $\pm$ 0.00028 & 1.38665 $\pm$ 3e-05 & 14000 $\pm$ 115 & 0.08179 $\pm$ 0.00038 & 1.24 $\pm$ 0.061 & 0.11832 $\pm$ 0.00049 & 0.155 $\pm$ 0.105 & $0.34\substack{+0.29 \\ -0.14}$ \\
77044471 & HATS-16 b & 1356.10219 $\pm$ 0.00142 & 2.68609 $\pm$ 0.00025 & 10861 $\pm$ 688 & 0.09089 $\pm$ 0.00255 & 1.132 $\pm$ 0.063 & 0.10422 $\pm$ 0.0033 & 0.345 $\pm$ 0.195 & $0.3\substack{+0.3 \\ -0.17}$ \\
77156657 & WASP-159 b & 1440.00942 $\pm$ 0.00413 & 3.84039 $\pm$ 0.00111 & 5501 $\pm$ 439 & 0.23246 $\pm$ 0.00759 & 1.471 $\pm$ 0.09 & 0.07417 $\pm$ 0.00296 & 0.389 $\pm$ 0.226 & $0.21\substack{+0.32 \\ -0.15}$ \\
78055054 & HATS-43 b & 1439.24967 $\pm$ 0.00069 & 4.38873 $\pm$ 0.00013 & 22247 $\pm$ 951 & 0.11018 $\pm$ 0.00211 & 1.274 $\pm$ 0.079 & 0.14916 $\pm$ 0.00319 & 0.259 $\pm$ 0.145 & $0.23\substack{+0.31 \\ -0.15}$ \\
92352620 & WASP-94 A b & 1328.29947 $\pm$ 0.0003 & 3.95005 $\pm$ 8e-05 & 11405 $\pm$ 167 & 0.17051 $\pm$ 0.00074 & 1.777 $\pm$ 0.079 & 0.10679 $\pm$ 0.00078 & 0.287 $\pm$ 0.078 & $0.35\substack{+0.29 \\ -0.14}$ \\
97409519 & WASP-124 b & 1327.0533 $\pm$ 0.00063 & 3.37284 $\pm$ 0.00015 & 15824 $\pm$ 954 & 0.09214 $\pm$ 0.00239 & 1.297 $\pm$ 0.076 & 0.1258 $\pm$ 0.00379 & 0.594 $\pm$ 0.107 & $0.23\substack{+0.32 \\ -0.16}$ \\
98283926 & HATS-42 b & 1494.56294 $\pm$ 0.00672 & 2.29198 $\pm$ 0.00125 & 9151 $\pm$ 2159 & 0.12184 $\pm$ 0.01199 & 1.178 $\pm$ --- & 0.09566 $\pm$ 0.01129 & 0.448 $\pm$ 0.261 & --- \\
98545929 & HATS-70 b & 1492.33656 $\pm$ 0.00476 & 1.88751 $\pm$ 0.00068 & 3670 $\pm$ 484 & 0.14413 $\pm$ 0.008 & 1.251 $\pm$ 0.108 & 0.06058 $\pm$ 0.00399 & 0.432 $\pm$ 0.252 & $0.26\substack{+0.31 \\ -0.17}$ \\
100100827 & WASP-18 b & 1354.45792 $\pm$ 5e-05 & 0.94145 $\pm$ 0.0 & 9629 $\pm$ 36 & 0.08143 $\pm$ 9e-05 & 1.286 $\pm$ 0.065 & 0.09813 $\pm$ 0.00018 & 0.375 $\pm$ 0.014 & $0.35\substack{+0.29 \\ -0.15}$ \\
102192004 & WASP-174 b & 1575.16346 $\pm$ 0.00541 & 4.2329 $\pm$ 0.00202 & 6049 $\pm$ 2514 & 0.06547 $\pm$ 0.01284 & 1.106 $\pm$ 0.237 & 0.07777 $\pm$ 0.01616 & 0.533 $\pm$ 0.301 & $0.62\substack{+0.21 \\ -0.39}$ \\
104024556 & WASP-167 b & 1571.96216 $\pm$ 0.00042 & 2.02203 $\pm$ 6e-05 & 9223 $\pm$ 234 & 0.10014 $\pm$ 0.00114 & 1.815 $\pm$ 0.082 & 0.09604 $\pm$ 0.00122 & 0.281 $\pm$ 0.171 & $0.6\substack{+0.19 \\ -0.12}$ \\
111991770 & WASP-15 b & 1603.1969 $\pm$ 0.00048 & 3.75209 $\pm$ 0.00014 & 8703 $\pm$ 269 & 0.13731 $\pm$ 0.00146 & 1.365 $\pm$ 0.06 & 0.09329 $\pm$ 0.00144 & 0.458 $\pm$ 0.117 & $0.33\substack{+0.29 \\ -0.15}$ \\
112099249 & WASP-184 b & 1604.17249 $\pm$ 0.0101 & 5.18235 $\pm$ 0.00548 & 7483 $\pm$ 2723 & 0.19287 $\pm$ 0.02452 & 1.386 $\pm$ 0.262 & 0.0865 $\pm$ 0.01574 & 0.475 $\pm$ 0.276 & $0.26\substack{+0.32 \\ -0.18}$ \\
112604564 & HATS-39 b & 1496.3127 $\pm$ 0.00687 & 4.57817 $\pm$ 0.00291 & 4271 $\pm$ 1078 & 0.13793 $\pm$ 0.01292 & 1.134 $\pm$ 0.178 & 0.06535 $\pm$ 0.00825 & 0.478 $\pm$ 0.274 & $0.44\substack{+0.27 \\ -0.27}$ \\
114299824 & HATS-55 b & 1493.60782 $\pm$ 0.00519 & 4.20423 $\pm$ 0.0017 & 8275 $\pm$ 3272 & 0.09267 $\pm$ 0.0123 & 0.944 $\pm$ 0.191 & 0.09097 $\pm$ 0.01798 & 0.488 $\pm$ 0.283 & $0.34\substack{+0.29 \\ -0.22}$ \\
114749636 & WASP-147 b & 1357.66508 $\pm$ 0.00832 & 4.60384 $\pm$ 0.00335 & 6780 $\pm$ 1653 & 0.16544 $\pm$ 0.01498 & 1.074 $\pm$ 0.141 & 0.08234 $\pm$ 0.01003 & 0.475 $\pm$ 0.27 & $0.24\substack{+0.33 \\ -0.17}$ \\
116156517 & WASP-190 b & 1357.37517 $\pm$ 0.01007 & 5.36625 $\pm$ 0.00403 & 6172 $\pm$ 1552 & 0.17539 $\pm$ 0.01959 & 1.312 $\pm$ 0.176 & 0.07856 $\pm$ 0.00987 & 0.472 $\pm$ 0.271 & $0.31\substack{+0.3 \\ -0.21}$ \\
117979897 & WASP-141 b & 1439.87082 $\pm$ 0.00115 & 3.31092 $\pm$ 0.00029 & 8370 $\pm$ 463 & 0.13877 $\pm$ 0.00283 & 1.203 $\pm$ 0.067 & 0.09149 $\pm$ 0.00253 & 0.36 $\pm$ 0.202 & $0.22\substack{+0.32 \\ -0.15}$ \\
118956453 & WASP-170 b & 1519.89228 $\pm$ 0.00438 & 2.34514 $\pm$ 0.00077 & 16207 $\pm$ 6507 & 0.07086 $\pm$ 0.01417 & 1.292 $\pm$ 0.268 & 0.12731 $\pm$ 0.02555 & 0.526 $\pm$ 0.298 & $0.39\substack{+0.28 \\ -0.25}$ \\
120610833 & WASP-45 b & 1354.77331 $\pm$ 0.00055 & 3.12609 $\pm$ 0.00011 & 12371 $\pm$ 1954 & 0.05405 $\pm$ 0.00381 & 0.947 $\pm$ 0.097 & 0.11123 $\pm$ 0.00878 & 0.775 $\pm$ 0.164 & $0.32\substack{+0.31 \\ -0.19}$ \\
122612091 & WASP-72 b & 1387.83226 $\pm$ 0.0006 & 2.21674 $\pm$ 4e-05 & 4123 $\pm$ 182 & 0.14677 $\pm$ 0.00179 & 1.662 $\pm$ 0.088 & 0.06421 $\pm$ 0.00142 & 0.612 $\pm$ 0.105 & $0.55\substack{+0.21 \\ -0.13}$ \\
124778445 & NGTS-3 A b & 1469.58297 $\pm$ 0.00242 & 1.67543 $\pm$ 0.00034 & 13739 $\pm$ 1390 & 0.0829 $\pm$ 0.00461 & 1.099 $\pm$ --- & 0.11721 $\pm$ 0.00593 & 0.363 $\pm$ 0.222 & --- \\
125739286 & NGTS-2 b & 1602.70634 $\pm$ 0.00832 & 4.51143 $\pm$ 0.0033 & 9661 $\pm$ 2460 & 0.17988 $\pm$ 0.01763 & 1.633 $\pm$ 0.224 & 0.09829 $\pm$ 0.01251 & 0.474 $\pm$ 0.266 & $0.27\substack{+0.32 \\ -0.18}$ \\
127530399 & WASP-132 b & 1602.85614 $\pm$ 0.00057 & 7.13349 $\pm$ 0.00041 & 15786 $\pm$ 618 & 0.11794 $\pm$ 0.00197 & 0.966 $\pm$ 0.072 & 0.12564 $\pm$ 0.00246 & 0.198 $\pm$ 0.133 & $0.3\substack{+0.3 \\ -0.17}$ \\
134537478 & KELT-14 b & 1493.27232 $\pm$ 0.00028 & 1.71 $\pm$ 4e-05 & 12424 $\pm$ 424 & 0.06787 $\pm$ 0.00163 & 1.636 $\pm$ 0.084 & 0.11146 $\pm$ 0.0019 & 0.841 $\pm$ 0.016 & $0.34\substack{+0.29 \\ -0.15}$ \\
139528693 & WASP-78 b & 1438.2005 $\pm$ 0.00089 & 2.17539 $\pm$ 0.00014 & 7444 $\pm$ 178 & 0.18165 $\pm$ 0.00162 & 1.732 $\pm$ 0.089 & 0.08628 $\pm$ 0.00103 & 0.205 $\pm$ 0.128 & $0.28\substack{+0.31 \\ -0.16}$ \\
139733308 & HATS-5 b & 1441.13394 $\pm$ 0.00072 & 4.76353 $\pm$ 0.00029 & 11763 $\pm$ 476 & 0.10892 $\pm$ 0.00177 & 0.924 $\pm$ 0.055 & 0.10846 $\pm$ 0.00219 & 0.306 $\pm$ 0.159 & $0.25\substack{+0.31 \\ -0.16}$ \\
144065872 & WASP-95 b & 1326.50593 $\pm$ 0.00014 & 2.18466 $\pm$ 2e-05 & 10488 $\pm$ 148 & 0.10666 $\pm$ 0.00051 & 1.234 $\pm$ 0.06 & 0.10241 $\pm$ 0.00072 & 0.42 $\pm$ 0.039 & $0.28\substack{+0.31 \\ -0.14}$ \\
145750719 & HATS-60 b & 1357.56289 $\pm$ 0.00725 & 3.56073 $\pm$ 0.0018 & 5432 $\pm$ 994 & 0.14999 $\pm$ 0.01134 & 1.053 $\pm$ 0.112 & 0.0737 $\pm$ 0.00674 & 0.445 $\pm$ 0.255 & $0.29\substack{+0.3 \\ -0.19}$ \\
149603524 & WASP-62 b & 1581.97114 $\pm$ 0.00013 & 4.41194 $\pm$ 1e-05 & 12465 $\pm$ 74 & 0.14082 $\pm$ 0.00027 & 1.319 $\pm$ 0.059 & 0.11165 $\pm$ 0.00033 & 0.283 $\pm$ 0.031 & $0.18\substack{+0.32 \\ -0.13}$ \\
152476657 & WASP-120 b & 1411.72876 $\pm$ 0.00056 & 3.61121 $\pm$ 7e-05 & 5330 $\pm$ 222 & 0.12947 $\pm$ 0.00171 & 1.197 $\pm$ 0.058 & 0.07301 $\pm$ 0.00152 & 0.644 $\pm$ 0.107 & $0.28\substack{+0.3 \\ -0.17}$ \\
157266693 & HATS-53 b & 1574.01852 $\pm$ 0.00645 & 3.85328 $\pm$ 0.00206 & 15884 $\pm$ 2852 & 0.1335 $\pm$ 0.01132 & 1.299 $\pm$ --- & 0.12603 $\pm$ 0.01132 & 0.42 $\pm$ 0.248 & --- \\
158623531 & WASP-105 b & 1355.73243 $\pm$ 0.00043 & 7.87285 $\pm$ 0.00012 & 12810 $\pm$ 246 & 0.13907 $\pm$ 0.00103 & 1.248 $\pm$ 0.072 & 0.11318 $\pm$ 0.00109 & 0.195 $\pm$ 0.105 & $0.32\substack{+0.3 \\ -0.13}$ \\
159951311 & WASP-139 b & 1387.57111 $\pm$ 0.00052 & 5.92443 $\pm$ 0.00011 & 10378 $\pm$ 363 & 0.10723 $\pm$ 0.00141 & 0.839 $\pm$ 0.056 & 0.10187 $\pm$ 0.00178 & 0.289 $\pm$ 0.147 & $0.33\substack{+0.3 \\ -0.17}$ \\
160148385 & WASP-96 b & 1354.31988 $\pm$ 0.00057 & 3.42527 $\pm$ 0.00014 & 14211 $\pm$ 466 & 0.08203 $\pm$ 0.00186 & 1.295 $\pm$ 0.072 & 0.11921 $\pm$ 0.00196 & 0.731 $\pm$ 0.041 & $0.32\substack{+0.3 \\ -0.15}$ \\
160578764 & WASP-192 b & 1601.27728 $\pm$ 0.00413 & 2.87866 $\pm$ 0.00096 & 7913 $\pm$ 2006 & 0.08735 $\pm$ 0.00889 & 1.167 $\pm$ 0.162 & 0.08896 $\pm$ 0.01127 & 0.462 $\pm$ 0.272 & $0.49\substack{+0.24 \\ -0.29}$ \\
160708862 & WASP-178 b & 1602.83593 $\pm$ 0.0059 & 3.3442 $\pm$ 0.00159 & 11104 $\pm$ 2255 & 0.13239 $\pm$ 0.01116 & 1.804 $\pm$ 0.195 & 0.10538 $\pm$ 0.0107 & 0.433 $\pm$ 0.253 & $0.28\substack{+0.31 \\ -0.18}$ \\
162922904 & WASP-171 b & 1575.33818 $\pm$ 0.00851 & 3.81929 $\pm$ 0.00275 & 4269 $\pm$ 951 & 0.1849 $\pm$ 0.01481 & 1.288 $\pm$ 0.158 & 0.06534 $\pm$ 0.00728 & 0.461 $\pm$ 0.264 & $0.34\substack{+0.29 \\ -0.22}$ \\
166833457 & WASP-98 b & 1413.16658 $\pm$ 0.00048 & 2.96249 $\pm$ 0.00011 & 27262 $\pm$ 934 & 0.06235 $\pm$ 0.00173 & 1.155 $\pm$ 0.061 & 0.16511 $\pm$ 0.00283 & 0.7 $\pm$ 0.036 & $0.16\substack{+0.32 \\ -0.12}$ \\
166836920 & WASP-99 b & 1387.96008 $\pm$ 0.00042 & 5.75255 $\pm$ 0.00021 & 4694 $\pm$ 79 & 0.20985 $\pm$ 0.00099 & 1.14 $\pm$ 0.053 & 0.06851 $\pm$ 0.00058 & 0.202 $\pm$ 0.118 & $0.28\substack{+0.3 \\ -0.15}$ \\
169226822 & WASP-127 b & 1548.1202 $\pm$ 0.00029 & 4.17813 $\pm$ 0.00011 & 10160 $\pm$ 167 & 0.16203 $\pm$ 0.00088 & 1.322 $\pm$ 0.059 & 0.1008 $\pm$ 0.00083 & 0.204 $\pm$ 0.104 & $0.21\substack{+0.32 \\ -0.14}$ \\
170102285 & WASP-23 b & 1470.66104 $\pm$ 0.00046 & 2.94445 $\pm$ 0.00013 & 18748 $\pm$ 816 & 0.08762 $\pm$ 0.00155 & 1.182 $\pm$ 0.082 & 0.13692 $\pm$ 0.00298 & 0.429 $\pm$ 0.123 & $0.3\substack{+0.3 \\ -0.17}$ \\
170634116 & WASP-79 b & 1412.89217 $\pm$ 0.0002 & 3.66238 $\pm$ 2e-05 & 11481 $\pm$ 130 & 0.13657 $\pm$ 0.00062 & 1.549 $\pm$ 0.076 & 0.10715 $\pm$ 0.00061 & 0.54 $\pm$ 0.02 & $0.17\substack{+0.32 \\ -0.13}$ \\
172598832 & HATS-40 b & 1470.76184 $\pm$ 0.00684 & 3.26701 $\pm$ 0.00195 & 4614 $\pm$ 397 & 0.211 $\pm$ 0.00838 & 1.377 $\pm$ 0.089 & 0.06793 $\pm$ 0.00292 & 0.382 $\pm$ 0.231 & $0.22\substack{+0.32 \\ -0.15}$ \\
176685457 & NGTS-9 b & 1518.95092 $\pm$ 0.00512 & 4.43434 $\pm$ 0.00166 & 5356 $\pm$ 1943 & 0.07501 $\pm$ 0.01124 & 0.984 $\pm$ 0.185 & 0.07318 $\pm$ 0.01327 & 0.509 $\pm$ 0.289 & $0.59\substack{+0.21 \\ -0.39}$ \\
178284730 & WASP-140 b & 1412.69733 $\pm$ 0.00014 & 2.23598 $\pm$ 1e-05 & 22441 $\pm$ 809 & 0.04385 $\pm$ 0.00127 & 1.231 $\pm$ 0.068 & 0.1498 $\pm$ 0.0027 & 0.845 $\pm$ 0.014 & $0.31\substack{+0.3 \\ -0.15}$ \\
178289267 & HATS-57 b & 1413.16241 $\pm$ 0.00275 & 2.35064 $\pm$ 0.00047 & 13396 $\pm$ 1926 & 0.09666 $\pm$ 0.00604 & 1.095 $\pm$ 0.097 & 0.11574 $\pm$ 0.00832 & 0.413 $\pm$ 0.238 & $0.22\substack{+0.32 \\ -0.16}$ \\
178367144 & WASP-180 A b & 1492.89794 $\pm$ 0.00498 & 3.40922 $\pm$ 0.00139 & 9968 $\pm$ 1716 & 0.12076 $\pm$ 0.00962 & 1.176 $\pm$ 0.114 & 0.09984 $\pm$ 0.0086 & 0.422 $\pm$ 0.254 & $0.25\substack{+0.31 \\ -0.17}$ \\
178879588 & HATS-63 b & 1439.38736 $\pm$ 0.00436 & 3.05584 $\pm$ 0.00106 & 12451 $\pm$ 3841 & 0.08538 $\pm$ 0.01037 & 1.128 $\pm$ -- & 0.11159 $\pm$ 0.01721 & 0.511 $\pm$ 0.28 & --- \\
179317684 & HD 271181 b & 1544.67047 $\pm$ 0.00041 & 4.23112 $\pm$ 2e-05 & 6828 $\pm$ 107 & 0.16654 $\pm$ 0.00086 & 1.338 $\pm$ 0.064 & 0.08263 $\pm$ 0.00065 & 0.222 $\pm$ 0.113 & $0.32\substack{+0.3 \\ -0.15}$ \\
183532609 & WASP-8 b & 1358.92009 $\pm$ 0.00038 & 8.1591 $\pm$ 0.00028 & 14219 $\pm$ 431 & 0.15755 $\pm$ 0.00189 & 1.156 $\pm$ 0.07 & 0.11925 $\pm$ 0.00181 & 0.601 $\pm$ 0.026 & $0.22\substack{+0.31 \\ -0.14}$ \\
183537452 & WASP-29 b & 1356.41475 $\pm$ 0.00026 & 3.92275 $\pm$ 7e-05 & 9510 $\pm$ 237 & 0.09915 $\pm$ 0.00086 & 0.603 $\pm$ 0.04 & 0.09752 $\pm$ 0.00122 & 0.201 $\pm$ 0.121 & $0.22\substack{+0.32 \\ -0.15}$ \\
184240683 & WASP-5 b & 1355.50812 $\pm$ 0.0003 & 1.62841 $\pm$ 3e-05 & 12372 $\pm$ 329 & 0.08855 $\pm$ 0.00086 & 1.184 $\pm$ 0.061 & 0.11123 $\pm$ 0.00148 & 0.369 $\pm$ 0.113 & $0.26\substack{+0.31 \\ -0.15}$ \\
192826603 & NGTS-1 b & 1438.08152 $\pm$ 0.00136 & 2.64723 $\pm$ 0.00014 & 25808 $\pm$ 10602 & 0.03305 $\pm$ 0.00688 & 0.888 $\pm$ 0.184 & 0.16065 $\pm$ 0.033 & 0.729 $\pm$ 0.245 & $0.41\substack{+0.28 \\ -0.2}$ \\
201604954 & HATS-29 b & 1657.82779 $\pm$ 0.00196 & 4.60897 $\pm$ 0.00165 & 14628 $\pm$ 2108 & 0.11993 $\pm$ 0.00439 & 1.311 $\pm$ 0.116 & 0.12095 $\pm$ 0.00871 & 0.434 $\pm$ 0.213 & $0.27\substack{+0.31 \\ -0.18}$ \\
204317710 & HATS-32 b & 1355.98236 $\pm$ 0.00219 & 2.8123 $\pm$ 0.0004 & 13641 $\pm$ 1003 & 0.10981 $\pm$ 0.00373 & 1.051 $\pm$ 0.06 & 0.11679 $\pm$ 0.00429 & 0.279 $\pm$ 0.178 & $0.18\substack{+0.32 \\ -0.13}$ \\
204376737 & WASP-6 b & 1357.39448 $\pm$ 0.00026 & 3.36103 $\pm$ 6e-05 & 20558 $\pm$ 579 & 0.09389 $\pm$ 0.00105 & 1.141 $\pm$ 0.059 & 0.14338 $\pm$ 0.00202 & 0.218 $\pm$ 0.113 & $0.22\substack{+0.32 \\ -0.15}$ \\
207077681 & WASP-144 b & 1326.05006 $\pm$ 0.00325 & 2.27867 $\pm$ 0.00051 & 13414 $\pm$ 4092 & 0.07952 $\pm$ 0.00878 & 1.366 $\pm$ 0.231 & 0.11582 $\pm$ 0.01767 & 0.595 $\pm$ 0.266 & $0.38\substack{+0.29 \\ -0.27}$ \\
211438925 & WASP-20 b & 1356.5656 $\pm$ 0.0004 & 4.89939 $\pm$ 0.00014 & 10137 $\pm$ 220 & 0.11765 $\pm$ 0.00102 & 1.176 $\pm$ 0.118 & 0.10068 $\pm$ 0.00109 & 0.694 $\pm$ 0.019 & $0.18\substack{+0.33 \\ -0.13}$ \\
219698950 & HATS-1 b & 1574.01109 $\pm$ 0.00048 & 3.44649 $\pm$ 0.00013 & 15418 $\pm$ 673 & 0.08535 $\pm$ 0.00177 & 1.103 $\pm$ 0.059 & 0.12417 $\pm$ 0.00271 & 0.637 $\pm$ 0.054 & $0.16\substack{+0.32 \\ -0.12}$ \\
229047362 & WASP-25 b & 1572.99066 $\pm$ 0.00026 & 3.76489 $\pm$ 7e-05 & 19779 $\pm$ 461 & 0.099 $\pm$ 0.00084 & 1.2 $\pm$ 0.055 & 0.14064 $\pm$ 0.00164 & 0.367 $\pm$ 0.082 & $0.19\substack{+0.32 \\ -0.13}$ \\
228381868 & WASP-165 b & 1356.67181 $\pm$ 0.00932 & 3.4658 $\pm$ 0.00236 & 5234 $\pm$ 1465 & 0.15513 $\pm$ 0.01564 & 1.251 $\pm$ 0.197 & 0.07235 $\pm$ 0.01013 & 0.49 $\pm$ 0.279 & $0.38\substack{+0.28 \\ -0.24}$ \\
230982885 & WASP-97 b & 1355.49043 $\pm$ 0.00014 & 2.07276 $\pm$ 2e-05 & 12064 $\pm$ 233 & 0.09616 $\pm$ 0.00063 & 1.206 $\pm$ 0.058 & 0.10984 $\pm$ 0.00106 & 0.415 $\pm$ 0.055 & $0.28\substack{+0.31 \\ -0.15}$ \\
231663901 & WASP-46 b & 1326.00912 $\pm$ 0.00038 & 1.43037 $\pm$ 4e-05 & 18830 $\pm$ 949 & 0.0556 $\pm$ 0.00139 & 1.189 $\pm$ 0.066 & 0.13722 $\pm$ 0.00346 & 0.649 $\pm$ 0.063 & $0.27\substack{+0.31 \\ -0.15}$ \\
231670397 & WASP-73 b & 1327.67385 $\pm$ 0.00077 & 4.0874 $\pm$ 0.00023 & 3263 $\pm$ 94 & 0.21821 $\pm$ 0.00134 & 1.233 $\pm$ 0.06 & 0.05713 $\pm$ 0.00083 & 0.337 $\pm$ 0.16 & $0.35\substack{+0.29 \\ -0.17}$ \\
232038798 & WASP-168 b & 1471.24991 $\pm$ 0.00323 & 4.15389 $\pm$ 0.00135 & 8692 $\pm$ 3941 & 0.05821 $\pm$ 0.01076 & 0.987 $\pm$ 0.228 & 0.09323 $\pm$ 0.02114 & 0.579 $\pm$ 0.307 & $0.47\substack{+0.29 \\ -0.35}$ \\
234112540 & CoRoT-5 b & 1470.41965 $\pm$ 0.00402 & 4.03792 $\pm$ 0.00159 & 16595 $\pm$ 5592 & 0.0941 $\pm$ 0.01128 & 1.34 $\pm$ 0.239 & 0.12882 $\pm$ 0.02171 & 0.486 $\pm$ 0.285 & $0.33\substack{+0.3 \\ -0.21}$ \\
238176110 & WASP-91 b & 1326.68919 $\pm$ 0.00026 & 2.79862 $\pm$ 5e-05 & 14612 $\pm$ 562 & 0.08649 $\pm$ 0.00117 & 1.066 $\pm$ 0.068 & 0.12088 $\pm$ 0.00232 & 0.446 $\pm$ 0.09 & $0.33\substack{+0.29 \\ -0.16}$ \\
248075138 & WASP-42 b & 1574.815 $\pm$ 0.00058 & 4.98151 $\pm$ 0.00023 & 15737 $\pm$ 756 & 0.10384 $\pm$ 0.00184 & 0.961 $\pm$ 0.061 & 0.12545 $\pm$ 0.00301 & 0.369 $\pm$ 0.151 & $0.18\substack{+0.32 \\ -0.13}$ \\
248111245 & HATS-27 b & 1573.47597 $\pm$ 0.00256 & 4.63683 $\pm$ 0.00039 & 8536 $\pm$ 516 & 0.18658 $\pm$ 0.00419 & 1.511 $\pm$ 0.113 & 0.09239 $\pm$ 0.00279 & 0.342 $\pm$ 0.19 & $0.22\substack{+0.32 \\ -0.16}$ \\
254113311 & TOI-1130 c & 1657.90545 $\pm$ 0.00747 & 8.35011 $\pm$ 0.00538 & 15227 $\pm$ 7249 & 0.05507 $\pm$ 0.01596 & 0.891 $\pm$ 0.227 & 0.1234 $\pm$ 0.02937 & 0.513 $\pm$ 0.296 & $0.63\substack{+0.21 \\ -0.35}$ \\
257567854 & WASP-22 b & 1411.90224 $\pm$ 0.00047 & 3.53265 $\pm$ 0.00011 & 10029 $\pm$ 247 & 0.12981 $\pm$ 0.00123 & 1.161 $\pm$ 0.051 & 0.10014 $\pm$ 0.00123 & 0.254 $\pm$ 0.13 & $0.2\substack{+0.32 \\ -0.14}$ \\
267263253 & HD 2685 b & 1325.78371 $\pm$ 0.0002 & 4.12692 $\pm$ 5e-05 & 9087 $\pm$ 78 & 0.1699 $\pm$ 0.00045 & 1.473 $\pm$ 0.06 & 0.09533 $\pm$ 0.00041 & 0.19 $\pm$ 0.081 & $0.22\substack{+0.32 \\ -0.14}$ \\
268644785 & KELT-15 b & 1494.13512 $\pm$ 0.00032 & 3.32945 $\pm$ 2e-05 & 9727 $\pm$ 164 & 0.1524 $\pm$ 0.00097 & 1.461 $\pm$ 0.063 & 0.09863 $\pm$ 0.00083 & 0.191 $\pm$ 0.11 & $0.26\substack{+0.31 \\ -0.16}$ \\
268766053 & WASP-53 b & 1386.23164 $\pm$ 0.00043 & 3.30969 $\pm$ 0.00012 & 17345 $\pm$ 862 & 0.07912 $\pm$ 0.00161 & 1.1 $\pm$ 0.073 & 0.1317 $\pm$ 0.00327 & 0.596 $\pm$ 0.071 & $0.3\substack{+0.3 \\ -0.17}$ \\
270468559 & HAT-P-42 b & 1519.46379 $\pm$ 0.00121 & 4.64135 $\pm$ 0.00056 & 6775 $\pm$ 359 & 0.15013 $\pm$ 0.00284 & 1.132 $\pm$ 0.063 & 0.08231 $\pm$ 0.00218 & 0.293 $\pm$ 0.184 & $0.34\substack{+0.29 \\ -0.17}$ \\
271168962 & WASP-131 b & 1601.5797 $\pm$ 0.00048 & 5.32218 $\pm$ 0.00019 & 6463 $\pm$ 171 & 0.13825 $\pm$ 0.00142 & 1.531 $\pm$ 0.079 & 0.08039 $\pm$ 0.00107 & 0.72 $\pm$ 0.031 & $0.49\substack{+0.23 \\ -0.13}$ \\
271893367 & TOI-150 b & 1548.86203 $\pm$ 0.00085 & 5.85742 $\pm$ 6e-05 & 5851 $\pm$ 176 & 0.22518 $\pm$ 0.00189 & 1.25 $\pm$ 0.058 & 0.07649 $\pm$ 0.00115 & 0.366 $\pm$ 0.137 & $0.16\substack{+0.32 \\ -0.12}$ \\
272212970 & HATS-67 b & 1572.50821 $\pm$ 0.00312 & 1.60909 $\pm$ 0.00037 & 12260 $\pm$ 4566 & 0.0587 $\pm$ 0.00831 & 1.538 $\pm$ --- & 0.11072 $\pm$ 0.02062 & 0.541 $\pm$ 0.286 & --- \\
272213425 & HATS-56 b & 1575.1164 $\pm$ 0.00874 & 4.32376 $\pm$ 0.00341 & 4672 $\pm$ 1183 & 0.17443 $\pm$ 0.01688 & 1.415 $\pm$ 0.192 & 0.06835 $\pm$ 0.00865 & 0.477 $\pm$ 0.273 & $0.4\substack{+0.27 \\ -0.25}$ \\
280210963 & CoRoT-12 b & 1471.02331 $\pm$ 0.00921 & 2.82763 $\pm$ 0.00239 & 56275 $\pm$ 14502 & 0.10253 $\pm$ 0.0139 & 2.258 $\pm$ 0.315 & 0.23722 $\pm$ 0.03057 & 0.398 $\pm$ 0.247 & $0.21\substack{+0.33 \\ -0.15}$ \\
281459670 & HATS-30 b & 1328.04041 $\pm$ 0.00041 & 3.17431 $\pm$ 4e-05 & 14075 $\pm$ 623 & 0.09938 $\pm$ 0.00166 & 1.336 $\pm$ 0.079 & 0.11864 $\pm$ 0.00262 & 0.451 $\pm$ 0.137 & $0.38\substack{+0.28 \\ -0.15}$ \\
281541555 & HATS-46 b & 1358.35617 $\pm$ 0.00154 & 4.74306 $\pm$ 0.00057 & 12590 $\pm$ 895 & 0.08719 $\pm$ 0.00293 & 0.899 $\pm$ 0.047 & 0.11221 $\pm$ 0.00399 & 0.329 $\pm$ 0.199 & $0.29\substack{+0.3 \\ -0.17}$ \\
281909674 & CoRoT-13 b & 1470.81292 $\pm$ 0.01306 & 4.03715 $\pm$ 0.00581 & 16980 $\pm$ 6235 & 0.11786 $\pm$ 0.02413 & 1.287 $\pm$ 0.252 & 0.13031 $\pm$ 0.02392 & 0.501 $\pm$ 0.279 & $0.27\substack{+0.33 \\ -0.2}$ \\
286865921 & WASP-83 b & 1573.61382 $\pm$ 0.00075 & 4.97109 $\pm$ 0.00031 & 10082 $\pm$ 412 & 0.12502 $\pm$ 0.00181 & 1.068 $\pm$ 0.062 & 0.10041 $\pm$ 0.00205 & 0.293 $\pm$ 0.156 & $0.33\substack{+0.29 \\ -0.16}$ \\
289793076 & HATS-13 b & 1328.08518 $\pm$ 0.00099 & 3.04386 $\pm$ 0.0002 & 21721 $\pm$ 1054 & 0.09518 $\pm$ 0.00209 & 1.086 $\pm$ 0.051 & 0.14738 $\pm$ 0.00358 & 0.297 $\pm$ 0.168 & $0.2\substack{+0.32 \\ -0.13}$ \\
290131778 & TOI-123 b & 1328.6837 $\pm$ 0.00039 & 3.30895 $\pm$ 9e-05 & 2992 $\pm$ 66 & 0.2194 $\pm$ 0.00099 & 1.397 $\pm$ 0.072 & 0.0547 $\pm$ 0.0006 & 0.561 $\pm$ 0.06 & $0.31\substack{+0.3 \\ -0.16}$ \\
294301883 & WASP-55 b & 1573.61482 $\pm$ 0.00101 & 4.46539 $\pm$ 0.0004 & 15567 $\pm$ 687 & 0.12967 $\pm$ 0.00248 & 1.324 $\pm$ 0.064 & 0.12477 $\pm$ 0.00275 & 0.265 $\pm$ 0.159 & $0.2\substack{+0.32 \\ -0.14}$ \\
306362738 & WASP-49 b & 1470.81972 $\pm$ 0.00037 & 2.78172 $\pm$ 0.0001 & 13218 $\pm$ 410 & 0.07114 $\pm$ 0.00157 & 1.142 $\pm$ 0.057 & 0.11497 $\pm$ 0.00178 & 0.759 $\pm$ 0.028 & $0.24\substack{+0.31 \\ -0.15}$ \\
308098254 & WASP-162 b & 1548.35224 $\pm$ 0.01104 & 9.62517 $\pm$ 0.00834 & 9323 $\pm$ 2820 & 0.16957 $\pm$ 0.02208 & 1.132 $\pm$ 0.181 & 0.09656 $\pm$ 0.0146 & 0.496 $\pm$ 0.283 & $0.31\substack{+0.3 \\ -0.21}$ \\
315002523 & HATS-26 b & 1518.61864 $\pm$ 0.00224 & 3.30257 $\pm$ 0.00022 & 8712 $\pm$ 452 & 0.19761 $\pm$ 0.00381 & 1.735 $\pm$ 0.101 & 0.09334 $\pm$ 0.00242 & 0.322 $\pm$ 0.182 & $0.22\substack{+0.32 \\ -0.15}$ \\
322307342 & HATS-68 b & 1328.48612 $\pm$ 0.00444 & 3.58495 $\pm$ 0.00108 & 5122 $\pm$ 938 & 0.12696 $\pm$ 0.00908 & 1.396 $\pm$ 0.148 & 0.07156 $\pm$ 0.00655 & 0.464 $\pm$ 0.261 & $0.58\substack{+0.19 \\ -0.3}$ \\
336732544 & HATS-62 b & 1327.97431 $\pm$ 0.00802 & 3.27725 $\pm$ 0.00215 & 15421 $\pm$ 5850 & 0.1068 $\pm$ 0.01984 & 1.141 $\pm$ --- & 0.12418 $\pm$ 0.02355 & 0.506 $\pm$ 0.283 & --- \\
336732616 & HATS-3 b & 1327.25298 $\pm$ 0.00059 & 3.54771 $\pm$ 0.00015 & 9338 $\pm$ 257 & 0.13219 $\pm$ 0.00139 & 1.2 $\pm$ 0.057 & 0.09664 $\pm$ 0.00133 & 0.302 $\pm$ 0.139 & $0.22\substack{+0.32 \\ -0.15}$ \\
339522221 & HATS-35 b & 1655.30248 $\pm$ 0.00465 & 1.821 $\pm$ 0.00055 & 11338 $\pm$ 1868 & 0.12074 $\pm$ 0.00843 & 1.402 $\pm$ 0.134 & 0.10648 $\pm$ 0.00877 & 0.425 $\pm$ 0.246 & $0.22\substack{+0.32 \\ -0.16}$ \\
355703913 & HATS-34 b & 1326.10628 $\pm$ 0.00111 & 2.10611 $\pm$ 7e-05 & 24728 $\pm$ 7403 & 0.03357 $\pm$ 0.01035 & 1.308 $\pm$ 0.205 & 0.15725 $\pm$ 0.02354 & 0.931 $\pm$ 0.051 & $0.21\substack{+0.32 \\ -0.13}$ \\
360742636 & HATS-33 b & 1654.16338 $\pm$ 0.00048 & 2.54955 $\pm$ 7e-05 & 13669 $\pm$ 554 & 0.09753 $\pm$ 0.00143 & 1.183 $\pm$ 0.064 & 0.11691 $\pm$ 0.00237 & 0.312 $\pm$ 0.145 & $0.28\substack{+0.31 \\ -0.15}$ \\
363851359 & HATS-54 b & 1601.77868 $\pm$ 0.00478 & 2.54422 $\pm$ 0.00101 & 6604 $\pm$ 1622 & 0.08449 $\pm$ 0.0088 & 1.035 $\pm$ --- & 0.08126 $\pm$ 0.00998 & 0.46 $\pm$ 0.264 & --- \\
379929661 & HAT-P-69 b & 1495.79529 $\pm$ 0.01002 & 4.7866 $\pm$ 0.00392 & 7482 $\pm$ 2092 & 0.19938 $\pm$ 0.02056 & 1.606 $\pm$ 0.232 & 0.0865 $\pm$ 0.0121 & 0.463 $\pm$ 0.27 & $0.23\substack{+0.33 \\ -0.17}$ \\
380589029 & HATS-24 b & 1653.97333 $\pm$ 0.00045 & 1.34851 $\pm$ 4e-05 & 17667 $\pm$ 554 & 0.0883 $\pm$ 0.00115 & 1.484 $\pm$ 0.083 & 0.13292 $\pm$ 0.00208 & 0.29 $\pm$ 0.134 & $0.23\substack{+0.32 \\ -0.15}$ \\
382391899 & WASP-50 b & 1411.09296 $\pm$ 0.0002 & 1.95512 $\pm$ 2e-05 & 19067 $\pm$ 633 & 0.06062 $\pm$ 0.00132 & 1.174 $\pm$ 0.059 & 0.13808 $\pm$ 0.00229 & 0.703 $\pm$ 0.029 & $0.22\substack{+0.32 \\ -0.14}$ \\
386259537 & WASP-169 b & 1521.89585 $\pm$ 0.01248 & 5.61189 $\pm$ 0.00663 & 5247 $\pm$ 1504 & 0.24902 $\pm$ 0.02639 & 1.56 $\pm$ 0.237 & 0.07243 $\pm$ 0.01038 & 0.483 $\pm$ 0.279 & $0.29\substack{+0.31 \\ -0.2}$ \\
388104525 & WASP-119 b & 1492.39663 $\pm$ 0.00068 & 2.49981 $\pm$ 0.00013 & 13138 $\pm$ 418 & 0.10818 $\pm$ 0.00145 & 1.269 $\pm$ 0.068 & 0.11462 $\pm$ 0.00182 & 0.219 $\pm$ 0.136 & $0.3\substack{+0.3 \\ -0.15}$ \\
393414358 & WASP-63 b & 1469.69711 $\pm$ 0.00058 & 4.37804 $\pm$ 0.00012 & 6051 $\pm$ 135 & 0.2084 $\pm$ 0.00153 & 1.357 $\pm$ 0.075 & 0.07779 $\pm$ 0.00087 & 0.188 $\pm$ 0.123 & $0.29\substack{+0.3 \\ -0.15}$ \\
398943781 & WASP-41 b & 1572.90615 $\pm$ 0.00024 & 3.05239 $\pm$ 6e-05 & 18582 $\pm$ 369 & 0.09599 $\pm$ 0.00078 & 1.167 $\pm$ 0.056 & 0.13632 $\pm$ 0.00135 & 0.176 $\pm$ 0.099 & $0.24\substack{+0.32 \\ -0.14}$ \\
399870368 & HAT-P-70 b & 1439.57484 $\pm$ 0.00706 & 2.74439 $\pm$ 0.00141 & 7590 $\pm$ 1838 & 0.13685 $\pm$ 0.01282 & 1.67 $\pm$ 0.21 & 0.08712 $\pm$ 0.01055 & 0.464 $\pm$ 0.267 & $0.3\substack{+0.3 \\ -0.2}$ \\
402026209 & WASP-4 b & 1355.18448 $\pm$ 0.00016 & 1.33823 $\pm$ 1e-05 & 23193 $\pm$ 405 & 0.07814 $\pm$ 0.00063 & 1.326 $\pm$ 0.083 & 0.15229 $\pm$ 0.00133 & 0.113 $\pm$ 0.078 & $0.21\substack{+0.32 \\ -0.15}$ \\
413376180 & HATS-2 b & 1572.12673 $\pm$ 0.00056 & 1.35416 $\pm$ 5e-05 & 17938 $\pm$ 724 & 0.07518 $\pm$ 0.00132 & 1.214 $\pm$ 0.081 & 0.13393 $\pm$ 0.0027 & 0.263 $\pm$ 0.147 & $0.32\substack{+0.3 \\ -0.17}$ \\
422655579 & WASP-71 b & 1413.14372 $\pm$ 0.00073 & 2.90344 $\pm$ 0.00014 & 4184 $\pm$ 112 & 0.19599 $\pm$ 0.00115 & 1.349 $\pm$ 0.067 & 0.06468 $\pm$ 0.00087 & 0.2 $\pm$ 0.131 & $0.35\substack{+0.29 \\ -0.15}$ \\
423275733 & WASP-142 b & 1518.69142 $\pm$ 0.00196 & 2.05301 $\pm$ 0.0003 & 9339 $\pm$ 991 & 0.09264 $\pm$ 0.00402 & 1.608 $\pm$ 0.127 & 0.09664 $\pm$ 0.00513 & 0.55 $\pm$ 0.228 & $0.52\substack{+0.22 \\ -0.23}$ \\
425206121 & KELT-19 A b & 1494.13554 $\pm$ 0.00033 & 4.61184 $\pm$ 0.00013 & 9311 $\pm$ 221 & 0.15711 $\pm$ 0.0012 & 1.571 $\pm$ 0.059 & 0.09649 $\pm$ 0.00115 & 0.367 $\pm$ 0.106 & $0.29\substack{+0.3 \\ -0.14}$ \\
437242640 & WASP-34 b & 1546.42188 $\pm$ 0.0003 & 4.31779 $\pm$ 0.0001 & 14396 $\pm$ 1408 & 0.0566 $\pm$ 0.00568 & 1.295 $\pm$ 0.098 & 0.11998 $\pm$ 0.00587 & 0.898 $\pm$ 0.052 & $0.28\substack{+0.31 \\ -0.16}$ \\
437248515 & WASP-31 b & 1547.48824 $\pm$ 0.00051 & 3.40591 $\pm$ 0.00014 & 15703 $\pm$ 428 & 0.08547 $\pm$ 0.0019 & 1.547 $\pm$ 0.069 & 0.12531 $\pm$ 0.00171 & 0.765 $\pm$ 0.029 & $0.27\substack{+0.31 \\ -0.15}$ \\
437261733 & HATS-59 b & 1548.79833 $\pm$ 0.00781 & 5.41703 $\pm$ 0.00429 & 12735 $\pm$ 3343 & 0.13168 $\pm$ 0.01492 & 1.203 $\pm$ --- & 0.11285 $\pm$ 0.01481 & 0.456 $\pm$ 0.266 & --- \\
437333618 & HATS-25 b & 1602.72421 $\pm$ 0.00141 & 4.2985 $\pm$ 0.00052 & 14333 $\pm$ 959 & 0.11388 $\pm$ 0.00333 & 1.303 $\pm$ 0.082 & 0.11972 $\pm$ 0.004 & 0.509 $\pm$ 0.18 & $0.25\substack{+0.31 \\ -0.18}$ \\
440777904 & HAT-P-24 b & 1491.69602 $\pm$ 0.0008 & 3.35521 $\pm$ 0.00017 & 9692 $\pm$ 288 & 0.13938 $\pm$ 0.00147 & 1.359 $\pm$ 0.07 & 0.09845 $\pm$ 0.00146 & 0.201 $\pm$ 0.127 & $0.3\substack{+0.3 \\ -0.16}$ \\
448589187 & WASP-175 b & 1547.69687 $\pm$ 0.005 & 3.06477 $\pm$ 0.00143 & 7485 $\pm$ 1505 & 0.1048 $\pm$ 0.00951 & 1.031 $\pm$ 0.116 & 0.08651 $\pm$ 0.0087 & 0.434 $\pm$ 0.253 & $0.33\substack{+0.29 \\ -0.21}$ \\
452808876 & WASP-82 b & 1438.96622 $\pm$ 0.00031 & 2.70583 $\pm$ 6e-05 & 5977 $\pm$ 63 & 0.19467 $\pm$ 0.00065 & 1.581 $\pm$ 0.085 & 0.07731 $\pm$ 0.00041 & 0.14 $\pm$ 0.09 & $0.3\substack{+0.31 \\ -0.16}$ \\
453789494 & WASP-172 b & 1604.28526 $\pm$ 0.01213 & 5.47675 $\pm$ 0.00617 & 8134 $\pm$ 2498 & 0.21449 $\pm$ 0.02548 & 1.836 $\pm$ 0.296 & 0.09019 $\pm$ 0.01385 & 0.503 $\pm$ 0.277 & $0.28\substack{+0.31 \\ -0.19}$ \\
455096220 & HAT-P-35 b & 1492.34126 $\pm$ 0.00169 & 3.64681 $\pm$ 0.00046 & 8675 $\pm$ 491 & 0.14924 $\pm$ 0.00333 & 1.23 $\pm$ 0.069 & 0.09314 $\pm$ 0.00263 & 0.307 $\pm$ 0.185 & $0.2\substack{+0.32 \\ -0.14}$ \\
455135327 & HAT-P-30 b & 1491.91574 $\pm$ 0.00037 & 2.81057 $\pm$ 8e-05 & 12016 $\pm$ 566 & 0.06812 $\pm$ 0.00243 & 1.429 $\pm$ 0.08 & 0.10962 $\pm$ 0.00258 & 0.852 $\pm$ 0.022 & $0.28\substack{+0.31 \\ -0.16}$ \\
466840711 & TOI-839 b & 1571.56494 $\pm$ 0.00644 & 2.48481 $\pm$ 0.00059 & 35823 $\pm$ 11966 & 0.08933 $\pm$ 0.01409 & 2.946 $\pm$ --- & 0.18927 $\pm$ 0.03161 & 0.44 $\pm$ 0.27 & --- \\
467971286 & HATS-69 b & 1656.6242 $\pm$ 0.00416 & 2.22496 $\pm$ 0.00063 & 11291 $\pm$ 2515 & 0.09274 $\pm$ 0.00853 & 0.913 $\pm$ --- & 0.10626 $\pm$ 0.01183 & 0.446 $\pm$ 0.264 & --- \\
468987719 & HAT-P-43 b & 1493.547 $\pm$ 0.00177 & 3.33343 $\pm$ 0.0005 & 13383 $\pm$ 865 & 0.11984 $\pm$ 0.0036 & 1.245 $\pm$ 0.068 & 0.11569 $\pm$ 0.00374 & 0.338 $\pm$ 0.191 & $0.22\substack{+0.32 \\ -0.15}$ \\
\enddata
    
\tablenotetext{4}{ExoFOP values were uploaded by Jason Eastman from global fits using EXOFAST \citep{eastman2013exofast}}

\end{deluxetable}
\end{longrotatetable}

\end{document}